\documentclass[a4paper, amsfonts, amssymb, amsmath, reprint, showkeys, nofootinbib, twoside,  superscriptaddress]{revtex4-1}

\usepackage[english]{babel}
\usepackage[utf8]{inputenc}
\usepackage[colorinlistoftodos, color=green!40, prependcaption]{todonotes}
\usepackage{amsthm}
\usepackage{mathtools}
\usepackage{physics}
\usepackage{xcolor}
\usepackage{graphicx}
\usepackage[left=23mm,right=13mm,top=35mm,columnsep=15pt]{geometry} 
\usepackage{adjustbox}
\usepackage{placeins}
\usepackage[T1]{fontenc}
\usepackage{lipsum}
\usepackage{csquotes}
\usepackage{graphicx}

\usepackage{siunitx}
\usepackage{float}
\usepackage[pdftex, pdftitle={Article}, pdfauthor={Author}]{hyperref} 
\begin{document}
\title{Integrated broadband optical isolator via dynamic rotating destructive interference}

\author{Kyunghun Han}
    \email{kyunghun.han@nist.gov}
    \affiliation{Microsystems and Nanotechnology Division, National Institute of
Standards and Technology, 100 Bureau Drive, Gaithersburg, Maryland, 20899,
USA}
    \affiliation{Theiss Research, La Jolla, California, 92037, USA}
    \affiliation{Whiting School of Engineering, Johns Hopkins University, 3400 N. Charles Street, Baltimore, Maryland, 21218, USA}
\author{Yiliang Bao}%
    \affiliation{Microsystems and Nanotechnology Division, National Institute of
Standards and Technology, 100 Bureau Drive, Gaithersburg, Maryland, 20899,
USA}
\author{Junyeob Song}%
    \affiliation{Microsystems and Nanotechnology Division, National Institute of
Standards and Technology, 100 Bureau Drive, Gaithersburg, Maryland, 20899,
USA}
    \affiliation{Theiss Research, La Jolla, California, 92037, USA}

\author{David Long}%
    \affiliation{Microsystems and Nanotechnology Division, National Institute of
Standards and Technology, 100 Bureau Drive, Gaithersburg, Maryland, 20899,
USA}
\author{Sean Bresler}%
    \affiliation{Microsystems and Nanotechnology Division, National Institute of
Standards and Technology, 100 Bureau Drive, Gaithersburg, Maryland, 20899,
USA}
    \affiliation{University of Maryland, College Park, Maryland, 20742, USA}
\author{Daron Westly}%
\author{Jason Gorman}%
\author{Thomas LeBrun}%
\author{Kartik Srinivasan}%
    \affiliation{Microsystems and Nanotechnology Division, National Institute of
Standards and Technology, 100 Bureau Drive, Gaithersburg, Maryland, 20899,
USA}

\author{Vladimir Aksyuk}%

   \email{vladimir.aksyuk@nist.gov}%
    \affiliation{Microsystems and Nanotechnology Division, National Institute of
Standards and Technology, 100 Bureau Drive, Gaithersburg, Maryland, 20899,
USA}

\date{\today} 

\begin{abstract}
Photonic integrated circuits route and shape light on a chip, but back-reflections feed back into coherent on-chip lasers, destabilizing operation and corrupting signals. Robust operation requires an integrated optical isolator that strongly suppresses backward propagation while maintaining low-loss, broadband forward transmission. However, prior on-chip isolators rely on magneto-optic materials or resonance-based filters, which respectively demand non-standard processes or inherently constrain bandwidth. Here, we propose and experimentally demonstrate a traveling-wave optical isolator without magnetic materials or resonant elements. By driving four parallel optical channels with  periodic RF waves, we realize dynamic rotating destructive interference that continuously cancels backward-propagating light while leaving forward-propagating light unaffected. We achieve about 30 dB isolation at a wavelength of 789.7 nm and maintain over 24 dB isolation across an approximately 30 nm bandwidth (770 nm to 800 nm), including >20 dB isolation for two simultaneous lasers within an approximately 10 nm wavelength window. This wavelength span covers key alkali atomic transitions, enabling strong suppression of feedback-induced frequency noise and laser instability in atomic spectroscopy \cite{Long2024Sub-DopplerLight}, laser cooling \cite{Isichenko2023PhotonicTrap}, and locking applications \cite{Martin2018CompactRubidium}. We demonstrate a practical, broadband on-chip isolator applicable from the visible to the near-infrared, which is a crucial step toward fully integrated photonic platforms. 

\end{abstract}

\maketitle

\section{Introduction} \label{sec:Introduction}

Photonic integrated circuits (PICs) bring complex optical functionalities on a single chip, enabling applications from precision spectroscopy \cite{Long2024Sub-DopplerLight,Yao2023IntegratedCircuits} to quantum sensing \cite{Loh2025OpticalLaser} and high-bandwidth communication \cite{Rizzo2023MassivelyLink}. Recently, ever more stable, low-noise and tunable laser sources have been integrated on PICs \cite{Snigirev2023UltrafastPhotonics,Siddharth2025UltrafastLaser}. This development highlights the need for integrated optical isolators for protecting such laser sources and ensuring signal integrity. Even small amounts of light reflecting back into a laser can introduce noise, cause its frequency to drift, or even shut down. Integrated optical isolators act as one-way barriers, blocking unwanted reflections before they reach the laser cavity. By eliminating feedback, they guarantee that photonic circuits on a chip deliver stable, reliable optical signals. Unlike electronic diodes, which exploit the simple nonlinear characteristic of a p-n junction, an optical isolator must rely on more complicated optical mechanisms to break the Lorentz reciprocity inherent in most optical media \cite{Jalas2013WhatIsolator}.

In general, Lorentz reciprocity can be broken in nonlinear optical systems \cite{Shi2015LimitationsReciprocity} or in linear optical systems coupled with external time-reversal symmetry breaking components through electro-optic, acousto-optic, and thermo-optic effects. An ideal optical isolator must strongly block backward-traveling light while providing no attenuation to forward-propagating light. To meet the stringent requirements of emerging PICs, integrated isolators should operate in the linear regime, provide strong isolation over a wide spectral range, and maintain their non-reciprocal behavior independent of optical power level. Traditionally, magneto-optical isolators \cite{Huang2017IntegratedRange,Yamaguchi2018Low-lossInput,Pintus2017Microring-BasedPhotonics,Zhang2019MonolithicPhotonics,Bi2011On-chipResonators,Yan2024Ultra-broadbandPlatform,Srinivasan2022ReviewPhotonics} based on the Faraday effect in magnetized materials satisfy these requirements by employing a static magnetic field to break time-reversal symmetry through a non-symmetric optical permittivity tensor \cite{Srinivasan2022ReviewPhotonics}. However, on-chip integration is challenging due to issues in complementary metal-oxide-semiconductor (CMOS) compatibility and large-scale heterogeneous integration of magneto-optical materials. This brings a motivation for magnet-free isolators with external time-dependent driving waves. Coupling optical signals to acoustic or radiofrequency (RF) traveling-waves (TW) provides an attractive alternative to achieve optical non-reciprocity through breaking time-reversal symmetry with spatiotemporal modulation in PICs. 

Traveling-wave isolators based on plasma dispersion effect \cite{Lira2012ElectricallyChip,Tzuang2014Non-reciprocalLight,Fang2012PhotonicModulation}, electro-optic effect \cite{Bhandare2005NovelMaterial,Doerr2011OpticalModulators,Dostart2021OpticalModulators,Doerr2014SiliconIsolator,Gao2024Thin-filmEtching,Dong2015Travelling-waveIsolators,Shah2023Visible-telecomNiobate} and acousto-optic effect \cite{Kittlaus2021ElectricallyPhotonics,Sohn2021ElectricallySplitting,Tian2021Magnetic-freeIsolator,Kittlaus2018Non-reciprocalModulation,Sohn2018Time-reversalCircuits} impose spatiotemporal refractive index changes along the waveguides (Extended Data Table 1). In the simplest case of a single mode waveguide under phase modulation, a driving RF wave travels opposite to the forward-propagating light but co-propagates with the backward-propagating light, thereby enabling carrier suppression of the backward-propagating light at a specific RF power. However, the suppressed carrier’s energy is transferred into harmonic sidebands that must be removed by external optical filters, adding insertion loss and system complexity. Unlike their magnetic counterparts, such single waveguide TW isolators \cite{Yu2023IntegratedNiobate} have a severely restricted isolation bandwidth making them impractical for broadband applications involving laser wavelength tuning or direct modulation, such as for spectroscopy and multi-channel optical communication. Another type of TW isolator exchanges optical power between two spatial modes within the same waveguide system through mode coupling \cite{Lira2012ElectricallyChip,Tzuang2014Non-reciprocalLight,Fang2012PhotonicModulation,Kittlaus2018Non-reciprocalModulation,Sohn2021ElectricallySplitting,Tian2021Magnetic-freeIsolator,Kittlaus2021ElectricallyPhotonics,Sohn2018Time-reversalCircuits}. This approach operates efficiently only at certain wavelengths where the two waveguide modes and the traveling-wave are frequency- and phase-matched, also limiting the isolation bandwidth. Previous work has explored using multiple waveguide modes in parallel single-mode waveguides, each modulated individually, to enhance performance. However, it is not currently known whether it is possible in principle to achieve complete and broadband traveling-wave isolation while leaving the forward-propagating light unaffected. All reported schemes still exhibit fundamentally unavoidable drawbacks, such as a substantial optical power remaining in unwanted modulation harmonics \cite{Doerr2014SiliconIsolator}, degraded isolation performance due to the fundamentally finite rise and fall times of the RF waveform \cite{Dong2015Travelling-waveIsolators}, or a significant transmission loss for the forward propagating light due to unavoidable modulation \cite{Bhandare2005NovelMaterial}.

Here, we propose and experimentally demonstrate an integrated traveling-wave broadband optical isolator achieving a forward-to-backward isolation of $\approx$ 30 dB at a fixed wavelength of 789.7 nm, broadband isolation exceeding 24 dB across a spectral bandwidth of 30 nm (770 nm to 800 nm; 14.6 THz), and simultaneous isolation for two optical sources within a 10 nm wavelength window exceeding 20 dB. The traveling-wave solution to integrated broadband isolation is implemented in the wavelength range particularly important for suppressing feedback-induced laser frequency noise and instability during spectroscopy or locking to alkali atomic transitions such as \textsuperscript{87}Rb and \textsuperscript{39}K (Fig. 1a). Using a 4-channel linear RF traveling-wave electro-optic phase modulator, our approach establishes the dynamic rotating destructive interference (DRDI), achieving continuous destructive interference under the time-periodic phase modulation. The DRDI scheme fully and continuously suppresses the backward optical power and therefore generates no detectable sidebands in the optical spectrum. More generally, because our isolator does not rely on any resonant structures or mode coupling, its design is wavelength-scalable from the visible to the telecommunication band, limited only by the bandwidth of passive optical components such as splitters and combiners. By overcoming the bandwidth and isolation trade-off of prior TW isolators, our work provides a practical solution to protect on-chip lasers and other reflection-sensitive photonic components in a fully integrated platform. 
   
\section{Result} \label{sec:Result}
 \begin{figure*}
     \centering
     \includegraphics[width=1\linewidth]{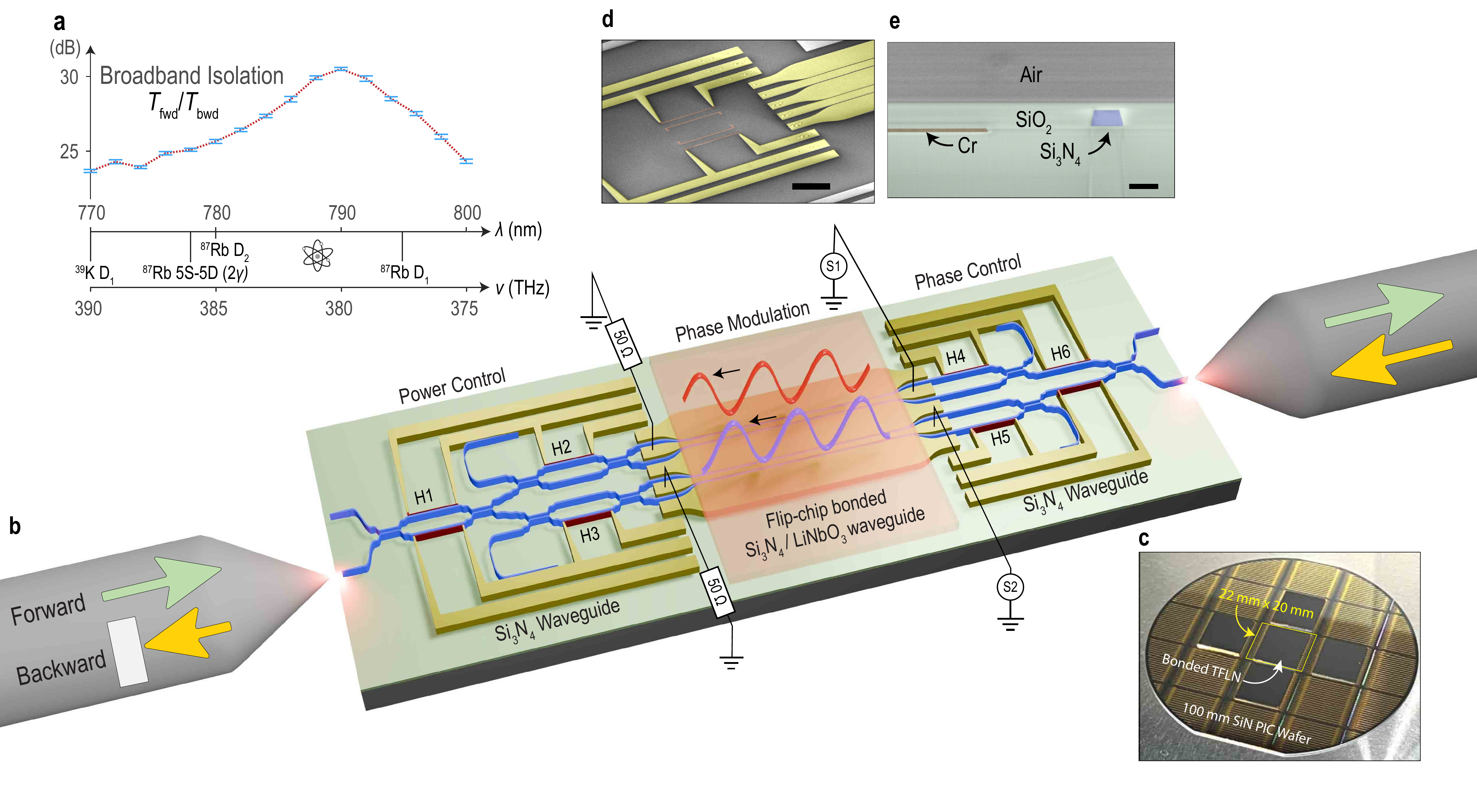}
     \caption{Integrated broadband traveling-wave optical isolator. \textbf{a}. Experimental optical isolation measured over the wavelength from 770 nm to 800 nm. This broadband isolator enables laser stability on multiple alkali atomic transitions such as those used in atomic clocks and laser cooling by suppressing unwanted back-reflections across this spectrum. \textbf{b.} Graphical illustration of the isolator, showing three functional regions: the left-side for power control (H1-H3) for balancing channel optical powers, the center for traveling-wave electro-optic phase modulation (red shaded region depicting the TFLN layer), the right-side for phase control (H4-H6). Blue colored waveguides form the optical path. Gold colored metal form the electrical paths for driving DC currents of integrated heaters (H1-H6) and RF waves (S1, S2) of the coplanar RF waveguides (terminated by 50~$\Omega$ loads). \textbf{c}. Photograph of a 100 mm diameter SiN PIC wafer after bonding TFLN pieces. The white-outlined chip, measuring 22 mm x 20 mm and containing 19 optical isolators, can be fabricated as nine separate dies on a single 100 mm diameter wafer. \textbf{d}. Bird’s-eye view false-colored SEM image of the region near H2 and H3 showing the gold RF and DC electrical lines (yellow) with Cr microheaters (red). The scale bar indicates 200 µm. \textbf{e}. False-colored SEM cross-section image of the SiN-only waveguide section showing SiN waveguide surrounded by SiO2 with Cr microheater (red) horizontally offset by 1.75 µm. The scale bar indicates 500 nm. }     \label{fig1}
 \end{figure*}
We implement the broadband isolator on a $\approx$ 350 nm thick silicon nitride photonic integrated circuit. The isolator uses a four-channel Mach-Zehnder modulator (MZM), providing four parallel interferometric paths (Figure 1b). The light with transverse-electric (TE) polarization enters the input waveguide and is split into these four channels by cascaded directional couplers. An X-cut  thin-film lithium niobate (TFLN) layer with a thickness of $\approx$ 150 nm is directly flip-chip bonded onto the passive Si\textsubscript{3}N\textsubscript{4} (SiN) waveguides to introduce electro-optic modulation, as indicated by the red shaded region in Fig. 1b. SiN/TFLN hybrid waveguide modes are phase-modulated by RF waves. These waves are applied via a gold coplanar RF waveguide (CPW) with a thickness of 900 nm buried beneath the TFLN (Extended Data Fig. 4). Two separate RF signals are applied to the two push-pull modulator pairs with the CPW lateral electrode gaps of $\approx$ 4 µm centered on the hybrid waveguides. Figure 1c shows a photograph of a 100 mm diameter SiN PIC wafer after TFLN bonding, demonstrating that nine 22 mm $\times$ 20 mm chips, containing 171 isolators in total, can be fabricated on a single wafer. Elsewhere, SiN-only waveguides surrounded by SiO\textsubscript{2} guide the light, and thermo-optic phase shifters are implemented using embedded Cr resistive microheaters (Figure 1d,e). The six quasi-static phase shifters (H1-H6) are arranged for accurate tuning of each channel’s optical power and phase. H1-H3 adjust the power balance through the cascaded directional coupler pairs \cite{Jin2019High-extinctionFilm}, and H4-H6 adjust the relative phase between the four modulated channels. When all four channels are set at equal amplitude and in-phase, forward transmission is maximized, and backward transmission is suppressed under the appropriate RF waveform. The resulting experimental optical isolation is shown in Figure 1a, giving $\geq $24 dB isolation over a $\approx$ 30 nm wavelength bandwidth. The isolation is defined in this paper as a ratio of time-averaged optical transmission in the forward direction (\textit{T}\textsubscript{fwd}) to that in the backward direction (\textit{T}\textsubscript{bwd}). Although the isolation principle applies generally for any wavelength, we chose the 770 nm to 800 nm range to support future integration for on-chip atomic spectroscopy \cite{Yang2007AtomicChip}, quantum sensing \cite{Schwindt2004Chip-scaleMagnetometer} and laser stabilization \cite{Hummon2018PhotonicInstability}, operating at key atomic transitions such as \textsuperscript{87}Rb D\textsubscript{2}, D\textsubscript{1} and two-photon lines. Suppression of back reflections is critical for precise wavelength tuning and stable frequency locking, and it is essential for integrated atomic clock and laser cooling systems that require precise stabilization of multiple lasers.
\begin{figure*}
    \centering
    \includegraphics[width=1\linewidth]{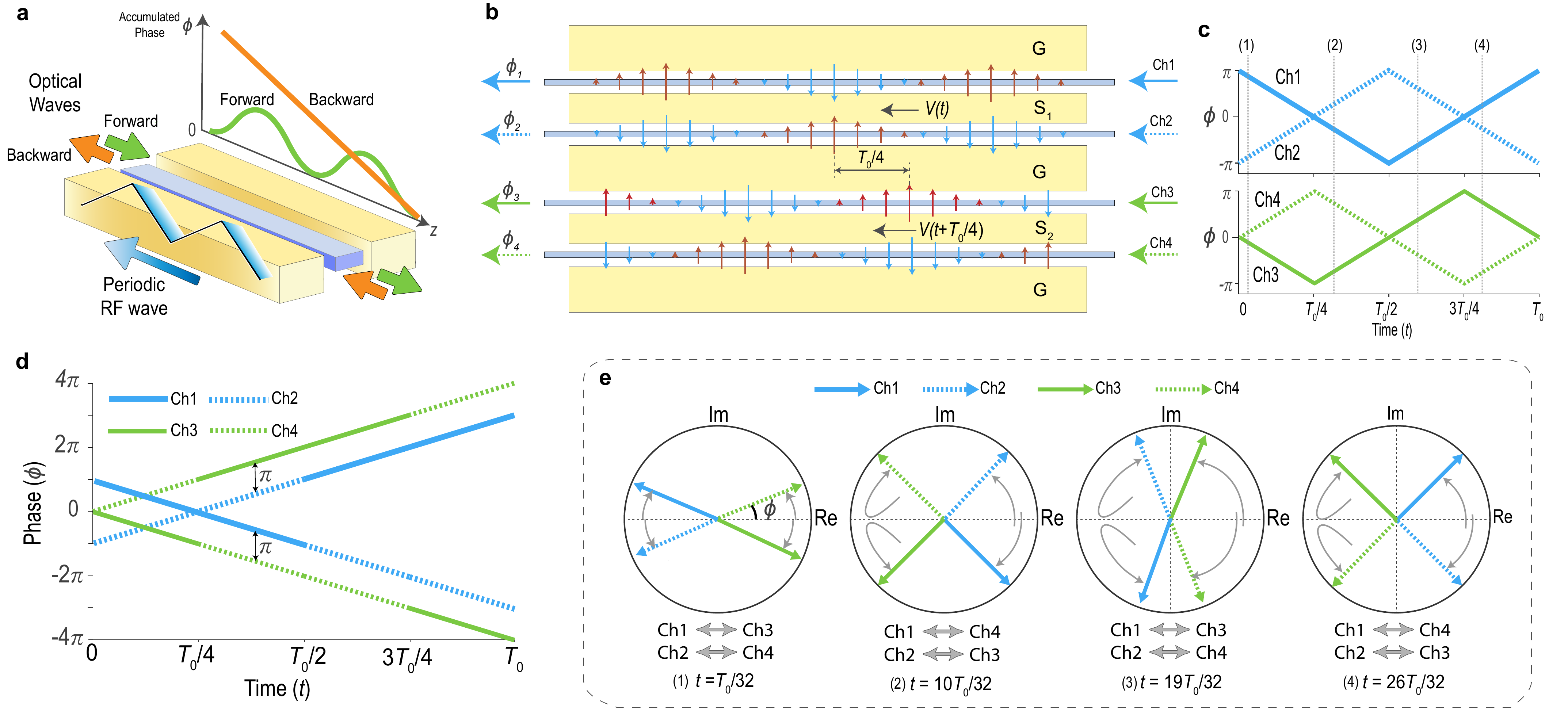}
 \caption{Dynamic rotating destructive interference with 4‑channel phase modulation.  
    \textbf{a}. Conceptual illustration of the single traveling‑wave phase‑modulator element. A periodic RF wave (blue triangle) co‑propagates with a backward‑propagating light (orange arrow) but counter‑propagates relative to a forward‑propagating light (green arrow). The plot on the side shows the accumulated phase ($\phi$) as a function of propagation distance for each direction, illustrating zero net phase shift for the forward‑propagating light and linear phase accumulation for the backward‑propagating light.  
    \textbf{b}. Top‑view schematic of the four waveguide channels (Ch1–4) with coplanar RF waveguides (G–S$_1$–G–S$_2$–G). Push–pull modulation is applied to Ch1 and Ch2 by S$_1 = V(t)$ and to Ch3 and Ch4 by S$_2 = V(t + T_0/4)$, a quarter‑period delayed waveform. Colored arrows between electrodes represent instantaneous electric‑field vectors.  
    \textbf{c}. Accumulated phases of the four channels over one RF period ($T_0$) with a peak phase modulation of $\pi$.  
    \textbf{d}. Time‑segmented, unwrapped phases from panel \textbf{c}. At any given time, two pairs of channels each maintain a $\pi$ phase difference and cancel each other. 
    \textbf{e}. Phasor diagrams at time instances ($t = T_0/32, 10T_0/32, 19T_0/32, 26T_0/32$), corresponding to vertical dashed lines in panel \textbf{c}, illustrating the rotating cancellation basis. Two channel vectors (one from the upper push–pull pair colored blue and one from the lower push–pull pair colored green) are opposite, canceling each other. These cancellation pairs rotate each quarter period, ensuring seamless destructive interference of the backward‑propagating light under dynamic phase modulation.}
  \label{fig:drdi}
\end{figure*}
For TW optical isolators, non-reciprocity arises from breaking the time-reversal symmetry of the optical medium through spatiotemporal modulation of the refractive index along the waveguides, induced by RF waves. When the RF wave propagates along an electro-optic phase modulator (Figure 2a), optical waves traveling in co-propagating and counter-propagating directions experience different accumulated phases, as described by,

\begin{equation}
\tag{1}
\phi_{\mathrm{FWD}}(t)
= \gamma \int_{0}^{L}
e^{-\frac{\alpha (L - z)}{2}}
V_{\mathrm{RF}}\bigl(t + \tfrac{z}{v_{\mathrm{opt}}} - \tfrac{L - z}{v_{\mathrm{RF}}}\bigr)
\,\mathrm{d}z
\end{equation}
\begin{equation}
\tag{2}
\phi_{\mathrm{BWD}}(t)
= \gamma \int_{0}^{L}
e^{-\frac{\alpha (L - z)}{2}}
V_{\mathrm{RF}}\bigl(t + \tfrac{L - z}{v_{\mathrm{opt}}} - \tfrac{L - z}{v_{\mathrm{RF}}}\bigr)
\,\mathrm{d}z
\end{equation}
where $z$ indicates the position along the modulator. $L$ indicates the modulation length. $\phi_{\mathrm{FWD}}$ denotes the accumulated phase for the forward‑propagating light, which counter‑propagates relative to the RF wave. For this case, the optical wave travels from $z=0$ to $z=L$ and the RF wave travels from $z=L$ to $z=0$. $\phi_{\mathrm{BWD}}$ represents the accumulated phase for the backward‑propagating light, which co‑propagates with the RF wave from $z=L$ to $z=0$. $V_{\mathrm{RF}}(t)$ refers to the instantaneous RF voltage waveform at a time $t$. $\alpha$ is the power attenuation constant of the RF wave, $v_{\mathrm{RF}}$ is the phase velocity of the RF wave, and $v_{\mathrm{opt}}$ is the group velocity of the light. $\gamma$ is the electro-optic phase modulation coefficient. When the forward‑propagating light interacts with a counter‑propagating RF wave with a zero‑mean periodic waveform, the instantaneous phase modulation experienced by the light is itself a zero‑mean periodic function. In the limit of negligible RF power attenuation ($\alpha \approx 0$), choosing
\begin{equation}
\tag{3}
f_{\mathrm{RF}}
= \frac{m}{\tfrac{L}{v_{\mathrm{RF}}} + \tfrac{L}{v_{\mathrm{opt}}}}
\end{equation}
where $m$ is a positive integer makes the total interaction time $t_{\mathrm{int}} = \tfrac{L}{v_{\mathrm{RF}}} + \tfrac{L}{v_{\mathrm{opt}}}$ equal to integer multiples of RF periods, resulting in zero net phase accumulation over the modulation length $L$. Under this condition, the forward-propagating light passes through the modulation region without experiencing any net phase shift due to the RF signal. While achieving this condition only requires adjusting the RF frequency, suppressing the backward-propagating light is far more challenging. This difficulty arises from the nature of phase modulation, which inherently generates an infinite series of harmonics. Achieving complete destructive interference for the backward light requires careful consideration of the phase relationships among all harmonic components with different phases. Identifying an optimal RF waveform consisting of multiple frequency components that leads to complete destructive interference is analytically nontrivial. This complexity can also arise because phase modulation is described by Bessel functions of the first kind, which, unlike polynomials, do not lead to simple closed-form solutions for harmonic cancellation. Rather than seeking a solution purely from a frequency domain perspective, it is more effective to consider the time domain perspective.

The fundamental principle for blocking coherent waves is destructive interference, which occurs when two or more waves combine so that their phasor vectors sum to zero. In practice, this can be achieved by combining waves of equal amplitude with constant phase difference over time. However, maintaining a constant phase difference between the coherent waves becomes challenging under periodic zero-mean phase modulation, which is required to keep the forward-propagating light unaffected. One simple possibility is to drive the phase modulation with two square waves or sawtooth waves with a half period delay to mimic the constant-phase condition using only two optical waves \cite{Dong2015Travelling-waveIsolators}. However, the abrupt waveform discontinuities inherent to these waveforms make this method impractical and significantly limit the achievable isolation performance (Extended Data Fig. 2). By contrast, triangular waveforms do not have waveform discontinuities in the time domain because they exhibit both positive and negative slopes, although discontinuities exist in their derivatives. Because triangular waveforms include both slope polarities, destructive interference between two phase modulated waves is impossible. There is always a crossing point at which their modulated phases coincide. We therefore developed a dynamic rotating destructive interference scheme to avoid this issue and achieve complete cancellation of phase modulated waves using four triangular or more generally, four time-delayed waveforms with both positive and negative slopes. Later in this paper, we introduce a waveform design procedure to remove derivative discontinuities inherent to triangular waveforms, while optimizing for higher isolation. Here, we use the triangular wave to describe the underlying principle. 
To implement the DRDI with the triangular wave, we use a four-channel electro-optic phase modulator with G-S$_1$-G-S$_2$-G coplanar RF waveguide driven by two RF waves (Figure 2b). Channel 1 (Ch1) and~2 (Ch2) form an upper push-pull pair, while channel~3 (Ch3) and~4 (Ch4) form a lower push-pull pair. Within a pair, two channels are driven by the same RF wave but opposite polarity (Extended Data Fig.  4). Furthermore, driving RF waves for the two push-pull pairs have time delay by one-quarter period ($T_0/4$) relative to each other. This configuration simplifies an experimental setup for equally spaced time delays of 0 (Ch1), $T_0/4$ (Ch3), $T_0/2$ (Ch2), and $3T_0/4$ (Ch4). When the phase modulation is driven by a triangular waveform with a peak-to-zero voltage amplitude at the half-wave voltage, $V_{\pi}$, the accumulated phase ($\phi$) across the four optical channels are shown in Figure 2c. To understand this mechanism more intuitively, we can time-segment and unwrap the accumulated phases ($\phi_1,\phi_2,\phi_3,\phi_4$) over time (Figure~2d). This reveals a structure in which, at any given moment, there are always two wave pairs that maintain a relative phase difference of $\pi$, ensuring continuous destructive interference. In phasor space, this corresponds to a dynamic rotation of cancellation pairs at every quarter-period, sustaining destructive interference throughout the RF wave period (Figure 2e). Ch1 is canceled by Ch3 during $0 \le t < T_0/4$, by Ch4 during $T_0/4 \le t < T_0/2$, and so on. Similarly, Ch2 alternates cancellation pair between Ch4 and Ch3 through the RF cycle. By rotating the destructive interference pairs in time, the overall waveform maintains a zero-mean condition, with both positive and negative slopes in the phase modulation, eliminating abrupt discontinuities found in the square and sawtooth waveforms. The triangular waveform represents a particular solution within a broader class of waveforms that satisfy the conditions for the DRDI. The general requirements for the driving RF waveform are further discussed in the Supplementary Information. In principle, DRDI can achieve arbitrarily high isolation under perfect amplitude and phase balances with sufficient harmonics. This represents a key advantage over resonance-based isolators which fundamentally must trade operational bandwidth for isolation. The DRDI imposes no such inherent compromises beyond the technological constraints of waveform generation and device uniformity.
\begin{figure*}
    \centering
    \includegraphics[width=1\linewidth]{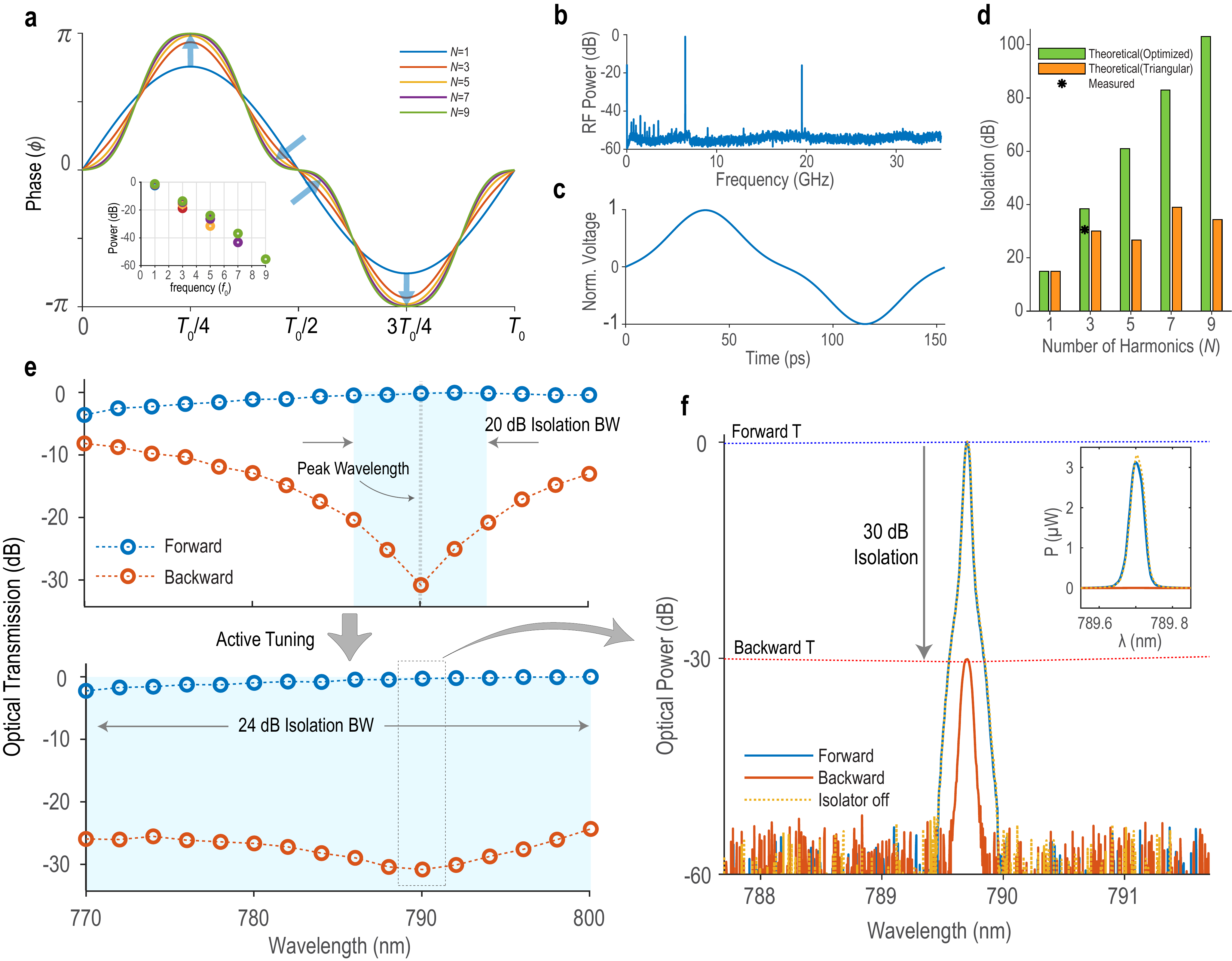}
 \caption{Broadband optical isolation demonstration with an optimized multi-harmonic RF waveform. \textbf{a}. Accumulated phase with optimized time-domain waveforms for a finite set of odd harmonics that satisfy half-wave symmetry. The inset shows the corresponding odd-harmonic power coefficients in a logarithmic scale. As the number of harmonics ($N$) increases, the peak-to-zero phase amplitude approaches $\pi$ and slopes at $t = 0, T_0/4, T_0/2$, and $3T_0/4$ approach zero to avoid modulated phase derivative difference to the quarter-period-delayed waveform. \textbf{b}. Measured RF power spectrum at the CPW output for the pre-distorted 6.5 GHz optimized waveform, with significant power only at the fundamental frequency and the 3rd harmonic. Power scale is normalized to the fundamental frequency power. \textbf{c}. Normalized voltage time-domain waveform reconstructed by inverse Fourier transform of the spectrum in panel \textbf{b}, showing rounded waveforms at maxima, minima, and zero-crossing points. Voltage scale is normalized to its maximum voltage. \textbf{d}. Theoretical isolation vs. number of harmonics for truncated harmonic triangular waveform (orange) and numerically optimized waveform (green). The experimentally measured isolation is shown as a black star. \textbf{e}. Wavelength dependency of normalized optical transmissions (normalized by active forward transmission at a wavelength of 800 nm) for the forward- (blue) and backward-propagating light (red) with (Upper) a fixed heater setting optimized at a wavelength of 789.7 nm and (Lower) an actively tuned heater setting for each wavelength. The blue shaded region represents the wavelength range where the isolations are above 20 dB. With actively tuned heaters for each wavelength, the isolation enhances >24 dB across 30 nm wavelength bandwidth. The measurement uncertainties, primarily caused by fiber-to-chip coupling drift, are estimated to be smaller than the plotted symbol size. \textbf{f}. Optical spectra measured near 789.7 nm with the isolator turned on for the forward- (blue) and backward-propagating optical wave (red), showing negligible sidebands. The spectrum of the forward-propagating light without the RF power (yellow) almost matches the forward spectrum with RF power. The optical powers are normalized to the peak power of “isolator off” spectrum. The time-averaged optical transmissions near this wavelength are plotted as dotted lines for the forward-propagating light (blue) and the backward-propagating light (red). The inset shows the same spectrum with linear scale optical powers. }
  \label{fig:EXPERIMENTAL}
\end{figure*}
The integrated microheaters are not strictly necessary for the proposed optical isolator with ideal static broadband 1x4 optical splitters and combiners. However, to compensate for experimental variations, we utilize thermo-optic effects in the SiN core and SiO$_2$ cladding to balance channel powers and set static phase offsets (Extended Data Fig. 5). The left-side heaters (Heater 1,2,3) balance the optical powers of all four optical channels, compensating for any residual splitting asymmetry introduced by the directional couplers as well as propagation loss differences. Heaters 2 and 4 tune the relative optical powers and optical phases of Ch1 and 2. Heaters 3 and 5 do the same for Ch3 and 4. Heaters 1 and 6 then adjust the relative optical powers and phases between the combined optical waves from the upper (Ch1 and 2) and lower (Ch3 and 4) push-pull pairs. By fixing Heater 1 such that only the upper pairs carry optical power, and sweeping Heater 2 and 4, we map the thermo-optic tuning response of that pair. The wavelength of the laser is centered at approximately 790 nm. At a DC power of 0.35 W, we achieve a full 2$\pi$ phase shift, showing two transmission minima when the optical powers of the channels are balanced. Even when the heaters are driven with 0.35 W of electrical power, no noticeable thermal drift is observed during measurement. Measured static extinction ratios are $\geq$ 32 dB (Ch1, 2), $\geq$ 30 dB (Ch 3, 4), and $\geq$ 31 dB for all four channels combined. These static extinction ratio measurements define the upper experimental limit of isolation performance for the electro-optic case, since both thermo-optic and electro-optic interference rely on the same linear optical behavior of the chip. In practice, electro-optic isolation may be further degraded by RF attenuation, velocity mismatch, finite electrode bandwidth, impedance mismatches, and parasitic reflections, so it cannot exceed the static thermo-optic bound. With optimized coupler designs, such as multimode-interference couplers \cite{Jin2023ParabolicPaths,Xie2022On-ChipInterferometer} and sub-wavelength assisted directional couplers \cite{Yun2018Ultra-broadbandWaveguides}, the static extinction ratio and bandwidth can be further improved. For the DRDI operation, all four channels must be combined in-phase at the output waveguide. Our protocol for finding the balanced powers and in-phase setting is described in the Supplementary Information. We characterize the electro-optic modulators by applying a low-frequency RF to the CPW while injecting light into one push-pull pair. A half-wave voltage ($V_\pi$) of a single-arm phase modulator is measured at $\approx$ 1.6 V, which is twice that of the push-pull case. The extinction ratio under low frequency RF closely matches the thermo-optic value, confirming that the thermo-optic case indeed represents the upper bound for electro-optic isolation performance.  

\begin{figure*}
    \centering
    \includegraphics[width=1\linewidth]{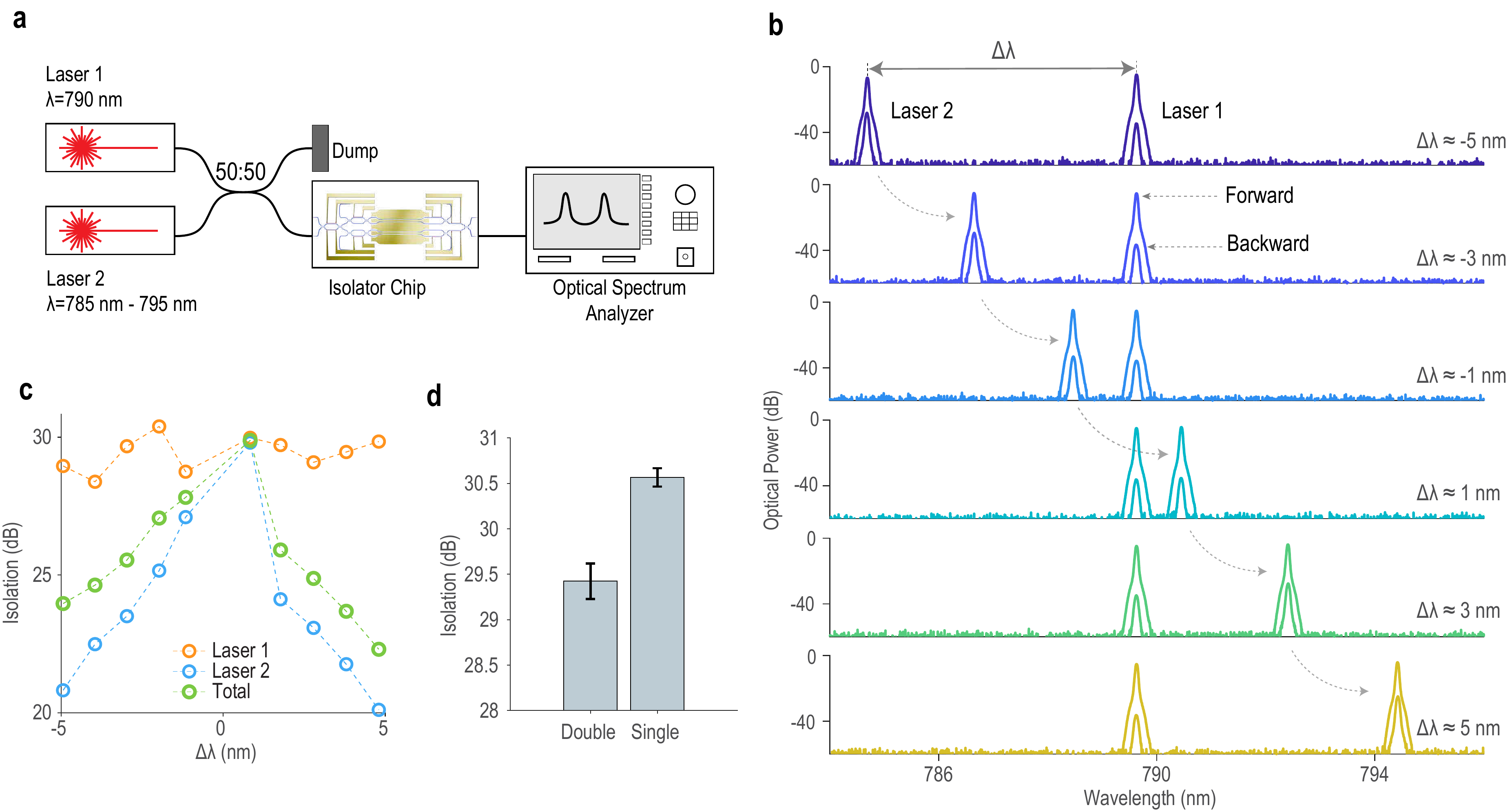}
 \caption{Double-laser experiment for broadband isolation performance. a. Experimental setup shows two external-cavity diode lasers (ECDLs), Laser 1 fixed at $\lambda_1$ $\approx$ 789.7 nm and Laser 2 swept through $\lambda_2$ $\approx$ 785 nm to 795 nm, are combined by a 50:50 optical fiber coupler. The combined signal passes through the isolator chip and is analyzed on an optical spectrum analyzer. For backward transmission, the input and output fibers are physically interchanged. The heaters are optimized for a wavelength near 790 nm. b. Forward- and backward-propagating optical spectra for various wavelength difference $\Delta\lambda$= $\lambda_1$- $\lambda_2$. In all cases, the backward transmissions for both lasers are suppressed by >20 dB, showing broadband isolation. The optical power scale is referenced to 1 mW. c. Extracted measured isolations from the optical spectra as a function of $\Delta\lambda$ for Laser 1 (orange circles), Laser 2 (blue circles), and the combined total signal (green circles). Laser 2 maintains >20 dB isolation across a 10 nm wavelength sweep, while the Laser 1 remains near 30 dB with minimal variation. d. Comparison of isolations at $\lambda$ $\approx$ 789.7 nm under single- and double-laser operation. Because of the additional laser, the isolation is reduced by $\approx$ 1.1 dB and the measurement uncertainty increases. Uncertainties are one standard deviation statistical uncertainties. }
  \label{fig:duallaser}
\end{figure*}
For a practical realization at a base frequency of several gigahertz, the ideal RF waveform must be approximated by truncating its Fourier series to a finite set of harmonics within the available RF waveform generation bandwidth. Without any adjustment, such truncation shifts the system away from the exact DRDI condition and reduces isolation. Figure 3d shows the theoretical isolation for truncated triangular waveforms with up to $N$ harmonics. The isolation remains below 40 dB, and for $N\leq$ 9, extra harmonics can even degrade performance. To maximize isolation for a given ${N}$, we use a gradient-descent algorithm to numerically optimize the amplitudes and phases of $N$ harmonics, while enforcing half-wave symmetry. Figure 3a plots the accumulated phases with these optimized waveforms for ${N}$ = 1, 3, 5, 7, and 9. As $N$ increases, the peak-to-zero phase amplitude ($\phi$) approaches $\pi$, and the slopes at $t$ = $T_0/4$ (peak), $T_0/2$ (zero-crossing), and $3T_0/4$ (dip) become gradually rounded, eliminating the derivative discontinuities of a triangular waveform. The rounded zero-crossing ensures that the derivatives of the modulated phase match that of the quarter-period-delayed waveform, which aligns at its peaks and dips. To maintain perfect cancellation of phase modulated light at those instants, the phase modulated waveforms must differ by $\pi$ and share identical instantaneous slopes to maintain perfect cancellation of the backward-propagating light. The resulting theoretical isolation increases monotonically with ${N}$ (Figure 3d). With only the first and third harmonics (${N}$ = 3), the optimized waveform achieves a theoretical isolation of 38 dB. This reduces the complexity and bandwidth requirements of the high frequency RF system. In practice, RF attenuation, velocity mismatch, finite CPW bandwidth and impedance mismatches will further limit the achievable isolation.
In the DRDI scheme, applying the correct RF waveform to the isolator is critical since most RF components introduce both linear and nonlinear distortions that scramble harmonic phases and amplitudes. The RF spectrum must contain only odd harmonics to satisfy the half-wave symmetry requirement of the DRDI, ${f(t+T_0/2) = -f(t)}$. We generate a 6.5 GHz RF waveform with an arbitrary waveform generator and then iteratively pre-distort it by monitoring the RF spectrum at the CPW output to compensate for distortions in the RF system. The frequency of 6.5 GHz is selected for the zero net accumulated phase for the forward-propagating light. This frequency exceeds the theoretical value predicted for the modulation length (${L}$) of 15 mm (Extended Data Fig. 3). We attribute this discrepancy to RF attenuation or a lower RF phase index arising from bonding artifacts such as air voids near the CPW. The optical transmission of the forward-propagating light remains unchanged by the driving RF (Extended Data Fig. 1). Any observed variation mostly falls within the measurement uncertainty. Due to RF component bandwidth limitations, we truncate the RF waveform at the third harmonic (19.5 GHz). The measured RF spectrum shows no even-order harmonics, and the odd-harmonic powers closely follow the theoretically optimized waveform (Figure 3b). The RF power measured at the output of the CPW is approximately 32 mW --- notably, our scheme does not require high watt level RF power or either optical or RF resonators. An inverse Fourier transform of that measured spectrum reconstructs a time-domain waveform that closely approximates the optimized waveform with rounded peaks, dips, and zero-crossing points (Figure 3c). With this optimized RF waveform, we measured spectra of forward- and backward-propagating lights by an optical spectrum analyzer (Figure 3f). The forward-to-backward peak-spectrum power ratio is measured at $\approx$ 30 dB, which directly approximates the isolation since no other sidebands exist in the spectra. Moreover, the spectrum of the forward-propagating light under RF is almost identical to the “Isolator off” case, which refers to the forward-propagating light with no RF, indicating no additional loss and frequency shift for the transmitted light. The frequency dependence of the forward transmission is discussed in Supplementary Information. To characterize wavelength dependency, we sweep the laser wavelength from 770 nm to 800 nm while holding the integrated heaters at their optimal in-phase, balanced power settings (tuned to 789.7 nm) (Figure 3e). The isolation varies approximately 21 dB across the 30 nm wavelength span, primarily due to the wavelength dependent splitting ratio of the parallel waveguide directional couplers. The maximum isolation is measured at 30.6 dB ± 0.1 dB. By actively tuning the heater powers for each wavelength, we achieve > 24 dB isolation over a 30 nm wavelength span. Replacing these directional couplers with broadband MMI or adiabatic designs would further flatten the response. To target other wavelength bands such as telecommunication C-band, one could re-design the power splitters and combiners for that wavelength while the DRDI principle itself remains unchanged.

To further demonstrate the distinctly broadband nature of our DRDI traveling-wave isolator, we rigorously evaluate the isolation performance under simultaneous multi-wavelength operation in a double-laser experiment (Figure 4a). Two external-cavity diode lasers were combined through a 50:50 fiber coupler, delivering equal optical power from both lasers to the isolator chip. Laser 1 was fixed at a wavelength of 789.7 nm, while Laser 2 was swept across the spectral range from nominally 785 nm to 795 nm in 1 nm increments. For each detuning $\Delta\lambda$= $\lambda_1$- $\lambda_2$, we measured forward and backward transmission spectra on an optical spectrum analyzer by interchanging the input and output fibers. The measured spectra are shown in Figure 4b. These results show that in every case the backward transmission of both lasers is suppressed by more than 20 dB across the entire 10 nm span. From each spectrum, we then numerically integrated the power spectral density over the peak region of each laser in both forward and backward directions to compute the isolation values. The resulting data are plotted in Figure 4c. As expected, the maximum total isolation occurs near 789.7 nm, corresponding to the heater settings optimized for that wavelength, and decays symmetrically as the wavelength is detuned. This roll-off closely matches the single-laser result discussed in Figure 3e. Small oscillations in Laser 1 isolation indicate minor perturbations introduced by Laser 2, despite the theoretical orthogonality of two different wavelengths at low optical power. Figure 4d presents a comparison of mean isolation at 789.7 nm under single-laser and double-laser operation. The double-laser case exhibits 29.4 dB $\pm$ 0.2 dB isolation, which is only approximately 1.1 dB reduction in mean isolation, accompanied by a slight increase in measurement uncertainty. These results confirm that our isolator reliably maintains more than 20 dB isolation over a 10 nm bandwidth for multiple lasers, with negligible crosstalk and without introducing any additional complexity into the RF electronics. These results represent a qualitative advance over earlier traveling-wave isolators, which relied on resonant features and were constrained by a narrow isolation bandwidth. The broadband isolation demonstrated here is essential for applications that require stable operation of multiple laser sources such as magneto-optical trapping of neutral atoms and optical frequency comb spectroscopy. Our approach removes resonant linewidth limitations and is not fundamentally cavity‑bandwidth‑limited, remaining constraints from splitter/combiner dispersion, finite waveform harmonics, and RF loss/mismatch.
\section{Conclusions} \label{sec:conclusions}
In summary, we have proposed and experimentally demonstrated the integrated broadband, traveling-wave optical isolator without magnetic materials or optical resonances. Using the four-channel Mach-Zehnder modulator driven by quarter-period delayed RF waveforms, we realize the dynamic rotating destructive interference that continuously suppresses the backward-propagating light while leaving the forward-propagating light unaffected. The isolator achieves approximately 30 dB isolation at 789.7 nm, and by leveraging on-chip microheater tuning we maintain above 24 dB isolation across a $\approx$ 30 nm wavelength bandwidth (770 nm to 800 nm). In the dual-laser experiment, the isolator simultaneously preserved high isolation (>20 dB) for two different wavelengths within a 10 nm wavelength window, confirming robust multi-wavelength performance. Importantly, the forward transmission showed no observable penalty such as additional loss due to the applied RF or unwanted sidebands, indicating key advantages over previous integrated isolators. This work shows a significant advance in non-reciprocal photonic devices, achieving combination of high isolation and broad bandwidth not seen in previous magnet-free integrated isolators. By breaking optical reciprocity through the DRDI scheme, our approach avoids the trade-off between isolation and bandwidth. In principle, the DRDI scheme imposes no fundamental limit on isolation aside from technological constraints, offering a route to even higher isolation with future waveguide directional coupler optimization or additional modulation harmonics. The platform is also extensible to other wavelength regimes. For example, with design modifications to splitters and combiners, the same principle could be applied in the telecommunication C-band or other infrared wavelengths, making it broadly useful. Practically, this isolator can be integrated with tunable lasers or frequency combs, eliminating the need of bulk isolators and enabling more complex and stable on-chip photonic systems. By solving a longstanding bottleneck in PIC design, the lack of a broadband on-chip isolator, our work paves the way for fully integrated photonic devices that can operate with one-way light flow for applications in data networks, quantum technology, and precision spectroscopy. 

\bibliographystyle{naturemag}  
\bibliography{references}    

\begin{thebibliography}{10}
\expandafter\ifx\csname url\endcsname\relax
  \def\url#1{\texttt{#1}}\fi
\expandafter\ifx\csname urlprefix\endcsname\relax\def\urlprefix{URL }\fi
\providecommand{\bibinfo}[2]{#2}
\providecommand{\eprint}[2][]{\url{#2}}

\bibitem{Long2024Sub-DopplerLight}
\bibinfo{author}{Long, D.~A.}, \bibinfo{author}{Stone, J.~R.}, \bibinfo{author}{Sun, Y.}, \bibinfo{author}{Westly, D.} \& \bibinfo{author}{Srinivasan, K.}
\newblock \bibinfo{title}{{Sub-Doppler spectroscopy of quantum systems through nanophotonic spectral translation of electro-optic light}}.
\newblock \emph{\bibinfo{journal}{Nature Photonics}} \textbf{\bibinfo{volume}{18}}, \bibinfo{pages}{1285--1292} (\bibinfo{year}{2024}).

\bibitem{Isichenko2023PhotonicTrap}
\bibinfo{author}{Isichenko, A.} \emph{et~al.}
\newblock \bibinfo{title}{{Photonic integrated beam delivery for a rubidium 3D magneto-optical trap}}.
\newblock \emph{\bibinfo{journal}{Nature Communications}} \textbf{\bibinfo{volume}{14}}, \bibinfo{pages}{3080} (\bibinfo{year}{2023}).

\bibitem{Martin2018CompactRubidium}
\bibinfo{author}{Martin, K.~W.} \emph{et~al.}
\newblock \bibinfo{title}{{Compact Optical Atomic Clock Based on a Two-Photon Transition in Rubidium}}.
\newblock \emph{\bibinfo{journal}{Physical Review Applied}} \textbf{\bibinfo{volume}{9}}, \bibinfo{pages}{014019} (\bibinfo{year}{2018}).

\bibitem{Yao2023IntegratedCircuits}
\bibinfo{author}{Yao, C.} \emph{et~al.}
\newblock \bibinfo{title}{{Integrated reconstructive spectrometer with programmable photonic circuits}}.
\newblock \emph{\bibinfo{journal}{Nature Communications}} \textbf{\bibinfo{volume}{14}}, \bibinfo{pages}{1--10} (\bibinfo{year}{2023}).

\bibitem{Loh2025OpticalLaser}
\bibinfo{author}{Loh, W.} \emph{et~al.}
\newblock \bibinfo{title}{{Optical atomic clock interrogation using an integrated spiral cavity laser}}.
\newblock \emph{\bibinfo{journal}{Nature Photonics}} \textbf{\bibinfo{volume}{19}}, \bibinfo{pages}{277--283} (\bibinfo{year}{2025}).

\bibitem{Rizzo2023MassivelyLink}
\bibinfo{author}{Rizzo, A.} \emph{et~al.}
\newblock \bibinfo{title}{{Massively scalable Kerr comb-driven silicon photonic link}}.
\newblock \emph{\bibinfo{journal}{Nature Photonics}} \textbf{\bibinfo{volume}{17}}, \bibinfo{pages}{781--790} (\bibinfo{year}{2023}).

\bibitem{Snigirev2023UltrafastPhotonics}
\bibinfo{author}{Snigirev, V.} \emph{et~al.}
\newblock \bibinfo{title}{{Ultrafast tunable lasers using lithium niobate integrated photonics}}.
\newblock \emph{\bibinfo{journal}{Nature}} \textbf{\bibinfo{volume}{615}}, \bibinfo{pages}{411--417} (\bibinfo{year}{2023}).

\bibitem{Siddharth2025UltrafastLaser}
\bibinfo{author}{Siddharth, A.} \emph{et~al.}
\newblock \bibinfo{title}{{Ultrafast tunable photonic-integrated extended-DBR Pockels laser}}.
\newblock \emph{\bibinfo{journal}{Nature Photonics}} \textbf{\bibinfo{volume}{19}}, \bibinfo{pages}{709--717} (\bibinfo{year}{2025}).

\bibitem{Jalas2013WhatIsolator}
\bibinfo{author}{Jalas, D.} \emph{et~al.}
\newblock \bibinfo{title}{{What is-and what is not-an optical isolator}}.
\newblock \emph{\bibinfo{journal}{Nature Photonics}} \textbf{\bibinfo{volume}{7}}, \bibinfo{pages}{579--582} (\bibinfo{year}{2013}).

\bibitem{Shi2015LimitationsReciprocity}
\bibinfo{author}{Shi, Y.}, \bibinfo{author}{Yu, Z.} \& \bibinfo{author}{Fan, S.}
\newblock \bibinfo{title}{{Limitations of nonlinear optical isolators due to dynamic reciprocity}}.
\newblock \emph{\bibinfo{journal}{Nature Photonics}} \textbf{\bibinfo{volume}{9}}, \bibinfo{pages}{388--392} (\bibinfo{year}{2015}).

\bibitem{Huang2017IntegratedRange}
\bibinfo{author}{Huang, D.} \emph{et~al.}
\newblock \bibinfo{title}{{Integrated broadband Ce:YIG/Si Mach–Zehnder optical isolators with over 100 nm tuning range}}.
\newblock \emph{\bibinfo{journal}{Optics Letters}} \textbf{\bibinfo{volume}{42}}, \bibinfo{pages}{4901} (\bibinfo{year}{2017}).

\bibitem{Yamaguchi2018Low-lossInput}
\bibinfo{author}{Yamaguchi, R.}, \bibinfo{author}{Shoji, Y.} \& \bibinfo{author}{Mizumoto, T.}
\newblock \bibinfo{title}{{Low-loss waveguide optical isolator with tapered mode converter and magneto-optical phase shifter for TE mode input}}.
\newblock \emph{\bibinfo{journal}{Optics Express}} \textbf{\bibinfo{volume}{26}}, \bibinfo{pages}{21271} (\bibinfo{year}{2018}).

\bibitem{Pintus2017Microring-BasedPhotonics}
\bibinfo{author}{Pintus, P.} \emph{et~al.}
\newblock \bibinfo{title}{{Microring-Based Optical Isolator and Circulator with Integrated Electromagnet for Silicon Photonics}}.
\newblock \emph{\bibinfo{journal}{Journal of Lightwave Technology}} \textbf{\bibinfo{volume}{35}}, \bibinfo{pages}{1429--1437} (\bibinfo{year}{2017}).

\bibitem{Zhang2019MonolithicPhotonics}
\bibinfo{author}{Zhang, Y.} \emph{et~al.}
\newblock \bibinfo{title}{{Monolithic integration of broadband optical isolators for polarization-diverse silicon photonics}}.
\newblock \emph{\bibinfo{journal}{Optica}} \textbf{\bibinfo{volume}{6}}, \bibinfo{pages}{473} (\bibinfo{year}{2019}).

\bibitem{Bi2011On-chipResonators}
\bibinfo{author}{Bi, L.} \emph{et~al.}
\newblock \bibinfo{title}{{On-chip optical isolation in monolithically integrated non-reciprocal optical resonators}}.
\newblock \emph{\bibinfo{journal}{Nature Photonics}} \textbf{\bibinfo{volume}{5}}, \bibinfo{pages}{758--762} (\bibinfo{year}{2011}).

\bibitem{Yan2024Ultra-broadbandPlatform}
\bibinfo{author}{Yan, W.} \emph{et~al.}
\newblock \bibinfo{title}{{Ultra-broadband magneto-optical isolators and circulators on a silicon nitride photonics platform}}.
\newblock \emph{\bibinfo{journal}{Optica}} \textbf{\bibinfo{volume}{11}}, \bibinfo{pages}{376} (\bibinfo{year}{2024}).

\bibitem{Srinivasan2022ReviewPhotonics}
\bibinfo{author}{Srinivasan, K.} \& \bibinfo{author}{Stadler, B. J.~H.}
\newblock \bibinfo{title}{{Review of integrated magneto-optical isolators with rare-earth iron garnets for polarization diverse and magnet-free isolation in silicon photonics}}.
\newblock \emph{\bibinfo{journal}{Optical Materials Express}} \textbf{\bibinfo{volume}{12}}, \bibinfo{pages}{697} (\bibinfo{year}{2022}).

\bibitem{Lira2012ElectricallyChip}
\bibinfo{author}{Lira, H.}, \bibinfo{author}{Yu, Z.}, \bibinfo{author}{Fan, S.} \& \bibinfo{author}{Lipson, M.}
\newblock \bibinfo{title}{{Electrically driven nonreciprocity induced by interband photonic transition on a silicon chip}}.
\newblock \emph{\bibinfo{journal}{Physical Review Letters}} \textbf{\bibinfo{volume}{109}}, \bibinfo{pages}{033901} (\bibinfo{year}{2012}).

\bibitem{Tzuang2014Non-reciprocalLight}
\bibinfo{author}{Tzuang, L.~D.}, \bibinfo{author}{Fang, K.}, \bibinfo{author}{Nussenzveig, P.}, \bibinfo{author}{Fan, S.} \& \bibinfo{author}{Lipson, M.}
\newblock \bibinfo{title}{{Non-reciprocal phase shift induced by an effective magnetic flux for light}}.
\newblock \emph{\bibinfo{journal}{Nature Photonics}} \textbf{\bibinfo{volume}{8}}, \bibinfo{pages}{701--705} (\bibinfo{year}{2014}).

\bibitem{Fang2012PhotonicModulation}
\bibinfo{author}{Fang, K.}, \bibinfo{author}{Yu, Z.} \& \bibinfo{author}{Fan, S.}
\newblock \bibinfo{title}{{Photonic Aharonov-Bohm effect based on dynamic modulation}}.
\newblock \emph{\bibinfo{journal}{Physical Review Letters}} \textbf{\bibinfo{volume}{108}}, \bibinfo{pages}{153901} (\bibinfo{year}{2012}).

\bibitem{Bhandare2005NovelMaterial}
\bibinfo{author}{Bhandare, S.} \emph{et~al.}
\newblock \bibinfo{title}{{Novel nonmagnetic 30-dB traveling-wave single-sideband optical isolator integrated in III/V material}}.
\newblock \emph{\bibinfo{journal}{IEEE Journal on Selected Topics in Quantum Electronics}} \textbf{\bibinfo{volume}{11}}, \bibinfo{pages}{417--421} (\bibinfo{year}{2005}).

\bibitem{Doerr2011OpticalModulators}
\bibinfo{author}{Doerr, C.~R.}, \bibinfo{author}{Dupuis, N.} \& \bibinfo{author}{Zhang, L.}
\newblock \bibinfo{title}{{Optical isolator using two tandem phase modulators}}.
\newblock \emph{\bibinfo{journal}{Optics Letters}} \textbf{\bibinfo{volume}{36}}, \bibinfo{pages}{4293--4295} (\bibinfo{year}{2011}).

\bibitem{Dostart2021OpticalModulators}
\bibinfo{author}{Dostart, N.}, \bibinfo{author}{Gevorgyan, H.}, \bibinfo{author}{Onural, D.} \& \bibinfo{author}{Popovi{\'{c}}, M.~A.}
\newblock \bibinfo{title}{{Optical isolation using microring modulators}}.
\newblock \emph{\bibinfo{journal}{Optics Letters}} \textbf{\bibinfo{volume}{46}}, \bibinfo{pages}{460} (\bibinfo{year}{2021}).

\bibitem{Doerr2014SiliconIsolator}
\bibinfo{author}{Doerr, C.~R.}, \bibinfo{author}{Chen, L.} \& \bibinfo{author}{Vermeulen, D.}
\newblock \bibinfo{title}{{Silicon photonics broadband modulation-based isolator}}.
\newblock \emph{\bibinfo{journal}{Optics Express}} \textbf{\bibinfo{volume}{22}}, \bibinfo{pages}{4493} (\bibinfo{year}{2014}).

\bibitem{Gao2024Thin-filmEtching}
\bibinfo{author}{Gao, L.} \emph{et~al.}
\newblock \bibinfo{title}{{Thin-film lithium niobate electro-optic isolator fabricated by photolithography assisted chemo-mechanical etching}}.
\newblock \emph{\bibinfo{journal}{Optics Letters}} \textbf{\bibinfo{volume}{49}}, \bibinfo{pages}{614} (\bibinfo{year}{2024}).

\bibitem{Dong2015Travelling-waveIsolators}
\bibinfo{author}{Dong, P.}
\newblock \bibinfo{title}{{Travelling-wave Mach-Zehnder modulators functioning as optical isolators}}.
\newblock \emph{\bibinfo{journal}{Optics Express}} \textbf{\bibinfo{volume}{23}}, \bibinfo{pages}{10498} (\bibinfo{year}{2015}).

\bibitem{Shah2023Visible-telecomNiobate}
\bibinfo{author}{Shah, M.}, \bibinfo{author}{Briggs, I.}, \bibinfo{author}{Chen, P.-K.}, \bibinfo{author}{Hou, S.} \& \bibinfo{author}{Fan, L.}
\newblock \bibinfo{title}{{Visible-telecom tunable dual-band optical isolator based on dynamic modulation in thin-film lithium niobate}}.
\newblock \emph{\bibinfo{journal}{Optics Letters}} \textbf{\bibinfo{volume}{48}}, \bibinfo{pages}{1978} (\bibinfo{year}{2023}).

\bibitem{Kittlaus2021ElectricallyPhotonics}
\bibinfo{author}{Kittlaus, E.~A.} \emph{et~al.}
\newblock \bibinfo{title}{{Electrically driven acousto-optics and broadband non-reciprocity in silicon photonics}}.
\newblock \emph{\bibinfo{journal}{Nature Photonics}} \textbf{\bibinfo{volume}{15}}, \bibinfo{pages}{43--52} (\bibinfo{year}{2021}).

\bibitem{Sohn2021ElectricallySplitting}
\bibinfo{author}{Sohn, D.~B.}, \bibinfo{author}{{\"{O}}rsel, O.~E.} \& \bibinfo{author}{Bahl, G.}
\newblock \bibinfo{title}{{Electrically driven optical isolation through phonon-mediated photonic Autler–Townes splitting}}.
\newblock \emph{\bibinfo{journal}{Nature Photonics}} \textbf{\bibinfo{volume}{15}}, \bibinfo{pages}{822--827} (\bibinfo{year}{2021}).

\bibitem{Tian2021Magnetic-freeIsolator}
\bibinfo{author}{Tian, H.} \emph{et~al.}
\newblock \bibinfo{title}{{Magnetic-free silicon nitride integrated optical isolator}}.
\newblock \emph{\bibinfo{journal}{Nature Photonics}} \textbf{\bibinfo{volume}{15}}, \bibinfo{pages}{828--836} (\bibinfo{year}{2021}).

\bibitem{Kittlaus2018Non-reciprocalModulation}
\bibinfo{author}{Kittlaus, E.~A.}, \bibinfo{author}{Otterstrom, N.~T.}, \bibinfo{author}{Kharel, P.}, \bibinfo{author}{Gertler, S.} \& \bibinfo{author}{Rakich, P.~T.}
\newblock \bibinfo{title}{{Non-reciprocal interband Brillouin modulation}}.
\newblock \emph{\bibinfo{journal}{Nature Photonics}} \textbf{\bibinfo{volume}{12}}, \bibinfo{pages}{613--619} (\bibinfo{year}{2018}).

\bibitem{Sohn2018Time-reversalCircuits}
\bibinfo{author}{Sohn, D.~B.}, \bibinfo{author}{Kim, S.} \& \bibinfo{author}{Bahl, G.}
\newblock \bibinfo{title}{{Time-reversal symmetry breaking with acoustic pumping of nanophotonic circuits}}.
\newblock \emph{\bibinfo{journal}{Nature Photonics}} \textbf{\bibinfo{volume}{12}}, \bibinfo{pages}{91--97} (\bibinfo{year}{2018}).

\bibitem{Yu2023IntegratedNiobate}
\bibinfo{author}{Yu, M.} \emph{et~al.}
\newblock \bibinfo{title}{{Integrated electro-optic isolator on thin-film lithium niobate}}.
\newblock \emph{\bibinfo{journal}{Nature Photonics}} \textbf{\bibinfo{volume}{17}}, \bibinfo{pages}{666--671} (\bibinfo{year}{2023}).

\bibitem{Jin2019High-extinctionFilm}
\bibinfo{author}{Jin, M.}, \bibinfo{author}{Chen, J.-Y.}, \bibinfo{author}{Sua, Y.~M.} \& \bibinfo{author}{Huang, Y.-P.}
\newblock \bibinfo{title}{{High-extinction electro-optic modulation on lithium niobate thin film}}.
\newblock \emph{\bibinfo{journal}{Optics Letters}} \textbf{\bibinfo{volume}{44}}, \bibinfo{pages}{1265} (\bibinfo{year}{2019}).

\bibitem{Yang2007AtomicChip}
\bibinfo{author}{Yang, W.} \emph{et~al.}
\newblock \bibinfo{title}{{Atomic spectroscopy on a chip}}.
\newblock \emph{\bibinfo{journal}{Nature Photonics}} \textbf{\bibinfo{volume}{1}}, \bibinfo{pages}{331--335} (\bibinfo{year}{2007}).

\bibitem{Schwindt2004Chip-scaleMagnetometer}
\bibinfo{author}{Schwindt, P.~D.} \emph{et~al.}
\newblock \bibinfo{title}{{Chip-scale atomic magnetometer}}.
\newblock \emph{\bibinfo{journal}{Applied Physics Letters}} \textbf{\bibinfo{volume}{85}}, \bibinfo{pages}{6409--6411} (\bibinfo{year}{2004}).

\bibitem{Hummon2018PhotonicInstability}
\bibinfo{author}{Hummon, M.~T.} \emph{et~al.}
\newblock \bibinfo{title}{{Photonic chip for laser stabilization to an atomic vapor with 10 -11 instability}}.
\newblock \emph{\bibinfo{journal}{Optica}} \textbf{\bibinfo{volume}{5}}, \bibinfo{pages}{443} (\bibinfo{year}{2018}).

\bibitem{Jin2023ParabolicPaths}
\bibinfo{author}{Jin, M.} \emph{et~al.}
\newblock \bibinfo{title}{{Parabolic MMI Coupler for 2 × 2 Silicon Optical Switch With Robustly High Extinction Ratio for Four Paths}}.
\newblock \emph{\bibinfo{journal}{IEEE Photonics Technology Letters}} \textbf{\bibinfo{volume}{35}}, \bibinfo{pages}{737--740} (\bibinfo{year}{2023}).

\bibitem{Xie2022On-ChipInterferometer}
\bibinfo{author}{Xie, S.}, \bibinfo{author}{Veilleux, S.} \& \bibinfo{author}{Dagenais, M.}
\newblock \bibinfo{title}{{On-Chip High Extinction Ratio Single-Stage Mach-Zehnder Interferometer Based on Multimode Interferometer}}.
\newblock \emph{\bibinfo{journal}{IEEE Photonics Journal}} \textbf{\bibinfo{volume}{14}}, \bibinfo{pages}{2237906} (\bibinfo{year}{2022}).

\bibitem{Yun2018Ultra-broadbandWaveguides}
\bibinfo{author}{Yun, H.}, \bibinfo{author}{Chrostowski, L.} \& \bibinfo{author}{Jaeger, N. A.~F.}
\newblock \bibinfo{title}{{Ultra-broadband 2 × 2 adiabatic 3 dB coupler using subwavelength-grating-assisted silicon-on-insulator strip waveguides}}.
\newblock \emph{\bibinfo{journal}{Optics Letters}} \textbf{\bibinfo{volume}{43}}, \bibinfo{pages}{1935} (\bibinfo{year}{2018}).

\bibitem{Fan2012AnDiode}
\bibinfo{author}{Fan, L.} \emph{et~al.}
\newblock \bibinfo{title}{{An All-Silicon Passive Optical Diode}}.
\newblock \emph{\bibinfo{journal}{Science}} \textbf{\bibinfo{volume}{335}}, \bibinfo{pages}{447--450} (\bibinfo{year}{2012}).

\bibitem{Abdelsalam2020LinearConversion}
\bibinfo{author}{Abdelsalam, K.}, \bibinfo{author}{Li, T.}, \bibinfo{author}{Khurgin, J.~B.} \& \bibinfo{author}{Fathpour, S.}
\newblock \bibinfo{title}{{Linear isolators using wavelength conversion}}.
\newblock \emph{\bibinfo{journal}{Optica}} \textbf{\bibinfo{volume}{7}}, \bibinfo{pages}{209} (\bibinfo{year}{2020}).

\bibitem{White2023IntegratedIsolators}
\bibinfo{author}{White, A.~D.} \emph{et~al.}
\newblock \bibinfo{title}{{Integrated passive nonlinear optical isolators}}.
\newblock \emph{\bibinfo{journal}{Nature Photonics}} \textbf{\bibinfo{volume}{17}}, \bibinfo{pages}{143--149} (\bibinfo{year}{2023}).

\bibitem{White2024UnifiedChip}
\bibinfo{author}{White, A.~D.} \emph{et~al.}
\newblock \bibinfo{title}{{Unified laser stabilization and isolation on a silicon chip}}.
\newblock \emph{\bibinfo{journal}{Nature Photonics}} \textbf{\bibinfo{volume}{18}}, \bibinfo{pages}{1305--1311} (\bibinfo{year}{2024}).

\end{thebibliography}
\clearpage
\clearpage
\section*{Methods}
\addcontentsline{toc}{section}{Methods}

\subsection*{Device fabrication}
The device was fabricated on a 100 mm diameter silicon wafer with a 3 µm wet thermal oxide layer and 350 nm stoichiometric silicon nitride (SiN). The SiN waveguides were patterned using an UV stepper lithography tool and a reactive-ion etcher (RIE). Firstly, the microheaters were patterned by the stepper, and then 100 nm of Chromium layer was deposited by electron-beam evaporator tool, followed by metal lift-off process with two-layer photoresists. SiO\textsubscript{2} cladding layer was deposited by plasma enhanced chemical vapor deposition and polished by chemical-mechanical- polish (CMP) leaving a 100 nm overcladding above the SiN waveguide. In the central area of the chip for the coplanar waveguide, RIE recessed the oxide layer until the underlying Cr was exposed, then continued for an additional 600 nm so that the gold layer lay below the top SiO2 surface, enabling direct bonding with the thin-film lithium niobate. The electrode area was patterned by the stepper, and then a 900 nm gold was deposited by electron-beam evaporator tool followed by metal lift-off process. A 4 inch, 300 nm lithium niobate-on-insulator (LNOI) wafer was thinned by CMP to a final thickness of 150 nm. Both wafers (SiN, LNOI) were coated with 5 nm of Al\textsubscript{2}O\textsubscript{3} via atomic layer deposition. The diced LNOI piece was flip-chip bonded onto the SiN wafer at room temperature and atmospheric pressure. The bonded wafer was annealed at 200 °C for 1 hour to increase the bonding strength. The wafer was diced into chips and their edges were polished by diamond polish pad.
\subsection*{Insertion loss}
The fiber-to-chip coupling loss is measured at 5.35 dB ± 0.30 dB per facet. The excess loss comes from the limited resolution of 365 nm photolithography tool, which cannot define the inverse taper tip narrowly enough. Device insertion loss is measured at 8.05 dB ± 0.28 dB at a nominal wavelength of 800 nm. We attribute the excessive insertion loss to imperfect fabrication during the bonding of the buried electrodes. Because this bonding process is not yet optimized, it may introduce additional scattering and increased optical propagation loss. However, the isolation principle can be applied to any traveling-wave modulator with low optical propagation loss.
\subsection*{PIC mode management}
To minimize mode mismatch loss between SiN only waveguide and SiN/TFLN hybrid waveguide, which arises from the higher refractive index of the TFLN, the SiN waveguide width is expanded to $\approx$ 2 µm near the edge of the bonding interface, confining the optical mode entirely within SiN and suppressing its evanescent overlap with TFLN. After the edge of bonding interface, the SiN width is adiabatically reduced to nominal 700 nm, transferring the mode into the TFLN for efficient electro-optic modulation. At the other end of the bonding interface, the SiN width is back to $\approx$ 2 µm for the same strategy. The redundant unused output port of the 2-by-2 directional coupler is routed adjacent to the buried gold electrode, which absorbs the residual light and prevents back-reflection.
\subsection*{Data analysis}
Wherever unspecified in the text, reported uncertainties are 68 percent confidence intervals, corresponding to one standard deviation of the mean. The statistical uncertainties are calculated from five measurements.
\section*{Acknowledgements} \label{sec:acknowledgements}

    National Institute of Standards and Technology NIST-on-a-Chip
Program. This work was performed under the financial assistance award 70NANB23H104 from the U.S. Department of Commerce, National Institute of Standards and Technology.
\section*{Author Contributions} \label{sec:Author Contribution}

K.H. conceived and developed the idea, designed the photonic circuits, participated in fabrication and performed all simulations and measurements; Y.B. led fabrication; J.S. assisted fabrication; D.L. and S.B. planned measurements; D.W. designed the fabrication process; J.G. and T.L. helped measurement planning and idea development; K.S. interpreted results and suggested the dual‑laser experiment; V.A. supervised the project and guided experiments and data analysis; K.H., K.S., and V.A. wrote the manuscript. All authors discussed the results and edited the manuscript.
\begin{table*}[htb!]
  \begin{flushleft}
  {Extended Data Table 1. Comparison of on-chip integrated optical isolators.}
  \end{flushleft}
  \label{tab:ED1}
  \begin{adjustbox}{width=\textwidth,center}
  \begin{tabular}{llllccccl}
    \hline
    Material      & Isolation Mechanism        & Device Structure                                         & Isolation‡ (dB) & Bandwidth† (nm)               & Wavelength (nm)  & Insertion Loss* (dB)                   & Ref        \\
    \hline
    SiN/TFLN      & EO/Spatiotemporal          & MZM                                                      & 30                   & 10                             & 790               & 8.05                                    & This work  \\
    Si            & TO                         & $\mu$‑ring resonator                                     & 28                   & $<0.05$                       & 1630              & 12                                      & \cite{Fan2012AnDiode}         \\
    Si            & Plasma/Temporal           & $\mu$‑ring                                    & 13.1                 & N/A                           & 1555              & 18.1                                    & \cite{Dostart2021OpticalModulators}         \\
    Si            & Plasma/Mode coupling       & Slotted waveguide                                        & 3                    & N/A                           & 1558              & 70                                      & \cite{Lira2012ElectricallyChip}         \\
    Si            & Plasma/Mode coupling       & Slotted waveguide                                        & 2.4                  & N/A                           & 1570              & Not explicitly reported                & \cite{Tzuang2014Non-reciprocalLight}         \\
    Si            & AO/Mode coupling           & Two multimode waveguide                                  & 38                   & 1                             & 1548              & 20**                                   & \cite{Kittlaus2018Non-reciprocalModulation}         \\
    AlN           & AO/Mode coupling           & Racetrack resonator                                      & 15                   & N/A                           & 1542              & Not explicitly reported                & \cite{Sohn2018Time-reversalCircuits}         \\
    Si/AlN        & AO/Mode coupling           & Single multimode waveguide                               & 16                   & N/A                           & 1524              & $>6$**                                 & \cite{Kittlaus2021ElectricallyPhotonics}        \\
    TFLN          & AO/Mode coupling           & Racetrack resonator                                      & 39.3                 & 0.00026                       & 1526              & 0.65                                    & \cite{Sohn2021ElectricallySplitting}         \\
    SiN/AlN       & AO/Mode coupling           & $\mu$‑ring resonator                                     & 10                   & N/A                           & 1546              & $<1$                                    & \cite{Tian2021Magnetic-freeIsolator}         \\
    Si/Ce:YIG     & MO/non-symmetric tensor    & Racetrack resonator                                      & 19.5                 & N/A                           & 1542              & 18.8                                    & \cite{Bi2011On-chipResonators}         \\
    Si/Ce:YIG     & MO/non-symmetric tensor    & $\mu$‑ring resonator                                     & 32                   & $<0.1$                       & 1552              & 2.3                                     & \cite{Pintus2017Microring-BasedPhotonics}         \\
    Si/Ce:YIG     & MO/non-symmetric tensor    & MZM                                                      & 29                   & 18                            & 1523              & 9                                       & \cite{Huang2017IntegratedRange}        \\
    Si/Ce:YIG     & MO/non-symmetric tensor    & Adiabatic taper                                          & 16                   & N/A                           & 1561              & 13                                      & \cite{Yamaguchi2018Low-lossInput}         \\
    SiN/Ce:YIG    & MO/non-symmetric tensor    & MZM                                                      & 28                   & 29                            & 1588              & 2.7                                     & \cite{Yan2024Ultra-broadbandPlatform}         \\
    TFLN          & Nonlinear/$\chi^{(2)}$     & Single waveguide                                         & 18                   & N/A                           & 1565              & 1                                       & \cite{Abdelsalam2020LinearConversion}         \\
    SiN           & Nonlinear/$\chi^{(3)}$     & $\mu$‑ring resonator                                     & 35                   & 0.00031                       & 1550              & 5.5                                     & \cite{White2023IntegratedIsolators}        \\
    SiN           & Nonlinear/$\chi^{(3)}$     & $\mu$‑ring resonator                                     & 14                   & N/A                           & 1550              & Not explicitly reported                & \cite{White2024UnifiedChip}         \\
    InP           & EO/Spatiotemporal          & Single waveguide                                         & $<6$                 & N/A                           & 1580              & 2.3                                     & \cite{Doerr2011OpticalModulators}         \\
    Bulk LN       & EO/Spatiotemporal          & MZM                                                      & 12.5                 & N/A                           & 1552              & 5.5                                     & \cite{Dong2015Travelling-waveIsolators}         \\
     TFLN          & EO/Spatiotemporal          & Single waveguide + $\mu$‑ring         & 34                   & $<0.002$ (filter linewidth)   & 1552              & 0.5                                     & \cite{Yu2023IntegratedNiobate}         \\
    TFLN          & EO/Temporal                & Single waveguide + $\mu$‑ring         & $<10$                & N/A                           & 775, 1550         & $<0.5$                                  &\cite{Shah2023Visible-telecomNiobate}         \\
      GaAs--AlGaAs  & EO/Spatiotemporal          & MZM                                                      & 30                   & 30                             & 1550              & 8*** (6 dB unavoidable loss)              & \cite{Bhandare2005NovelMaterial}         \\
    \hline
  \end{tabular}
  \end{adjustbox}
  \begin{flushleft}
    ‡ Isolation(dB) = 10 log\textsubscript{10}($T_{\mathrm{fwd}}/T_{\mathrm{bwd}}$)\\
    † Bandwidth for isolation $>$ 20 dB without active tuning\\
    * Without fiber-to-chip coupling loss\\
    ** Maximum modulation efficiency to sideband\\
    *** Forward-propagating light is modulated by single-sideband modulation, which causes fundamentally unavoidable 6 dB loss\\
    AO: Acousto-optic effect, TO: Thermo-optic effect, EO: Electro-optic effect, Plasma: Plasma dispersion effect
  \end{flushleft}
\end{table*}

\begin{figure*}
    \centering
    \includegraphics[width=1\linewidth]{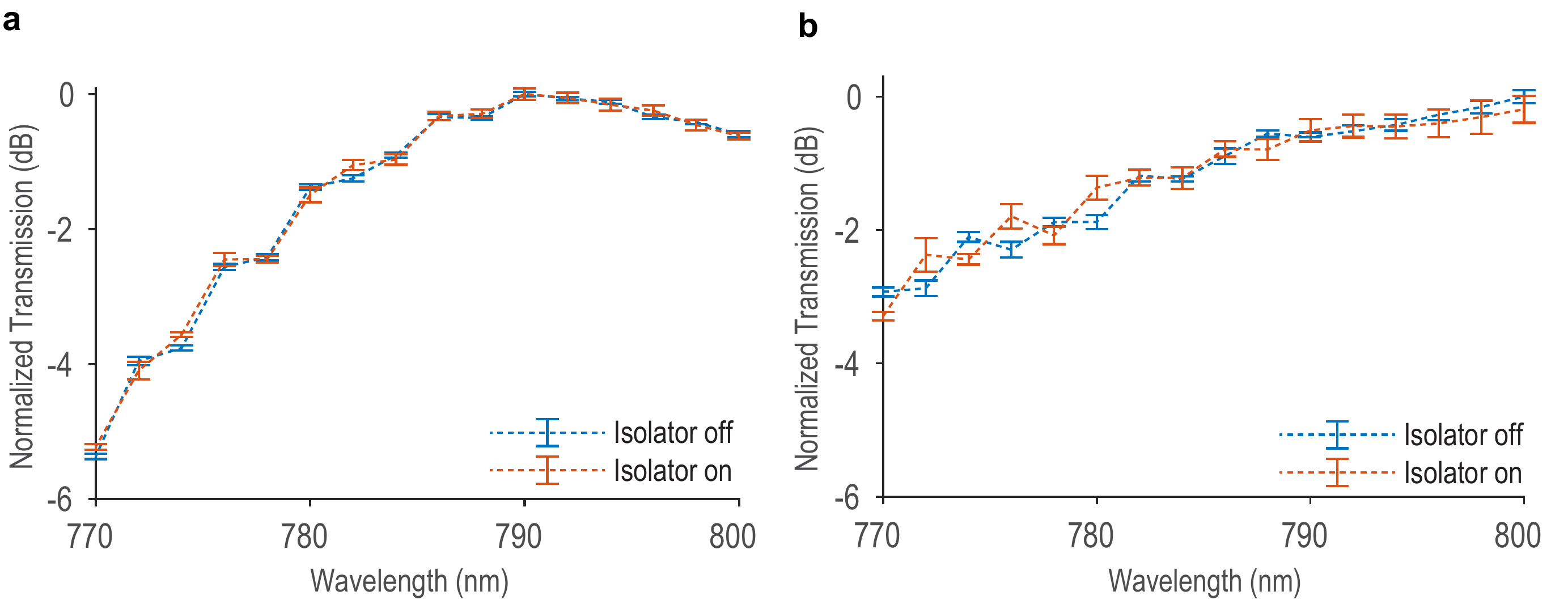}
      \begin{flushleft}
Extended Data Figure 1. Effect of a 6.5 GHz RF drive on the forward-propagating transmission. Normalized forward transmission is plotted as a function of wavelength from 770 nm to 800 nm for \textbf{a}. a fixed optimization wavelength at 789.7 nm and \textbf{b}. actively tuned and optimized for every wavelength. The blue dotted line and its error bars represent the transmission with no RF, while the orange dotted line and its error bars show the transmission under the 6.5 GHz RF wave. For the fixed optimization case, both lines have a peak transmission at nominal 790 nm which is the optimized wavelength for the microheaters, and the transmission values are normalized to this peak. For the active tuning case, a peak transmission occurs at nominal 800 nm, and the transmission values are normalized to this peak. The two lines overlap at most wavelengths, indicating that the RF wave imposes no additional insertion loss across the wavelength range.
    \end{flushleft}
  \label{fig:EXF1}
\end{figure*}

\begin{figure*}
    \centering
    \includegraphics[width=1\linewidth]{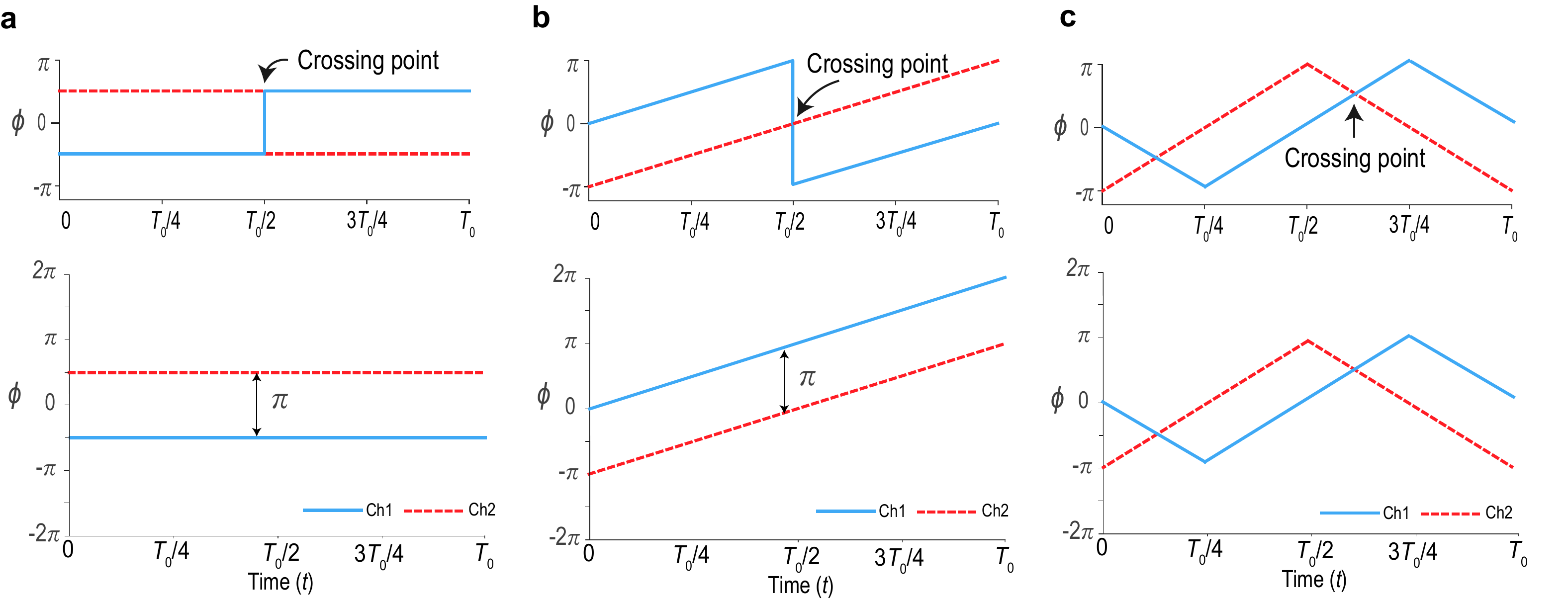}
          \begin{flushleft}
          Extended Data Figure 2. Accumulated phase for two time-delayed RF drive waveforms. \textbf{a}-\textbf{c}. Accumulated optical phase (top row) and the same phase traces segmented and aligned in time (bottom row) over one RF period for three waveform shapes are presented. \textbf{a}. Square waveform gives an ideal constant $\pi$ phase difference over each half period, resulting in continuous destructive interference but requires infinite transition slopes at every half periods. Finite transition slopes introduce partial destructive interference and reduce isolation performance at the phase crossing points. \textbf{b}. Sawtooth wave produces linear phase ramps but requires infinite transition slopes as well, leading to the same limitation in isolation performance. \textbf{c}. Triangular waveform has both positive and negative slopes which can eliminate the infinite transition slope but produce the unavoidable phase crossing points where two phases become equal, preventing continuous destructive interference throughout the RF period. The DRDI can overcome this limitation by introducing two additional RF waves and periodically rotating the canceling pairs so that each pair maintains a continuous destructive interference. 
        \end{flushleft}
  \label{fig:EXF2}
\end{figure*}

\begin{figure*}
    \centering
    \includegraphics[width=1\linewidth]{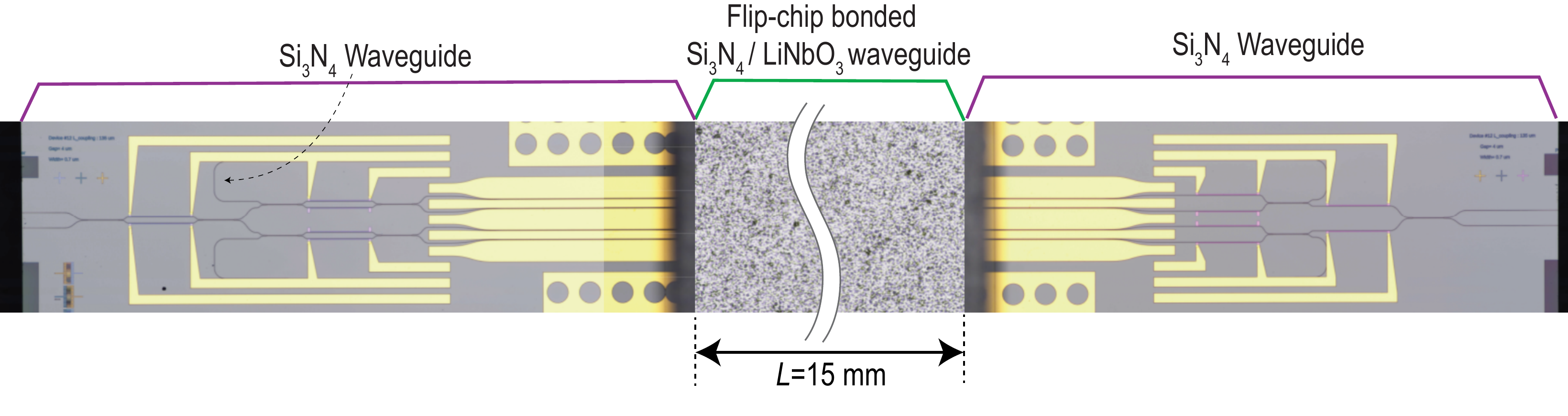}
              \begin{flushleft}
          Extended Data Figure 3. Stitched optical microscopy image of the fabricated isolator. Black narrow horizontally spanning lines indicate silicon nitride optical waveguides. Adjacent chromium microheaters tune the power balance and phase differences among the four parallel channels in the chip’s central region, where flip-chip bonded lithium niobate couples to the silicon nitride waveguide to form an electro-optically active hybrid waveguide. The coplanar RF waveguide carrying the RF driving waves modulates each channel’s optical phase over a length of approximately 15 mm.
        \end{flushleft}
  \label{fig:EXF3}
\end{figure*}

\begin{figure*}[p]
    \centering
    \includegraphics[width=1\linewidth]{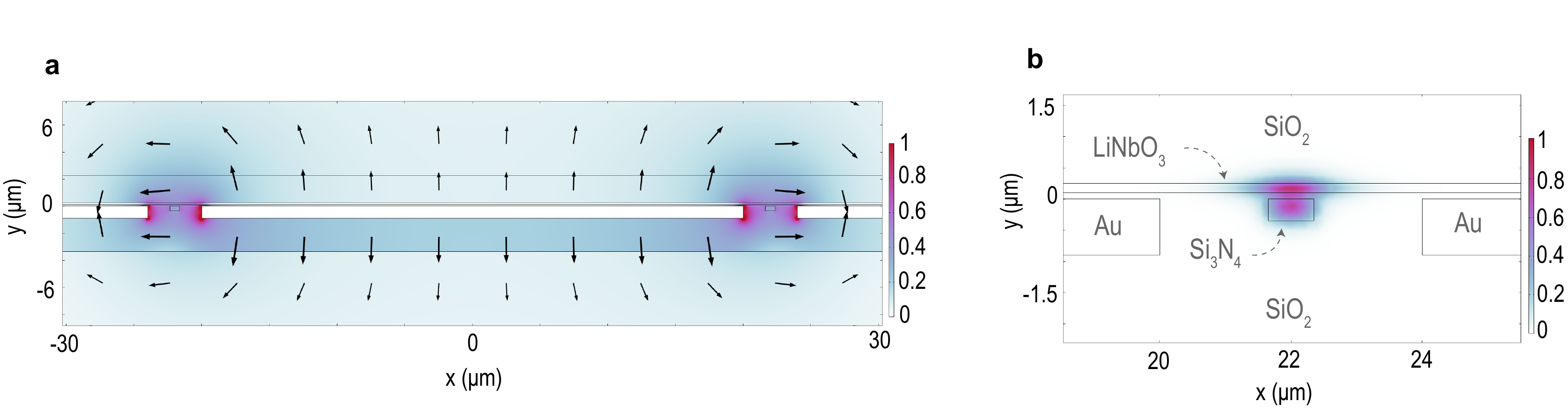}
              \begin{flushleft}
          Extended Data Figure 4. Electrostatic field and optical mode profile at the center electro-optic modulation area. \textbf{a}. Simulated electric field amplitude and vector distribution (black arrows) for a push-pull coplanar RF waveguide. The center buried electrode serves as a signal line, with adjacent ground lines. The electric field is concentrated at the electrode gap edges. The arrow directions indicate opposite electric field polarities between the push-pull pair channels. \textbf{b}. Simulated optical mode profile shows that the transverse-electric mode is confined within the silicon nitride waveguide and thin-film lithium niobate layer surrounded by the silicon dioxide cladding, providing strong overlap in lithium niobate for efficient electro-optic modulation while keeping metal layers outside the optical mode to avoid absorption loss.
        \end{flushleft}
  \label{fig:EXF4}
\end{figure*}

\begin{figure*}
    \centering
    \includegraphics[width=1\linewidth]{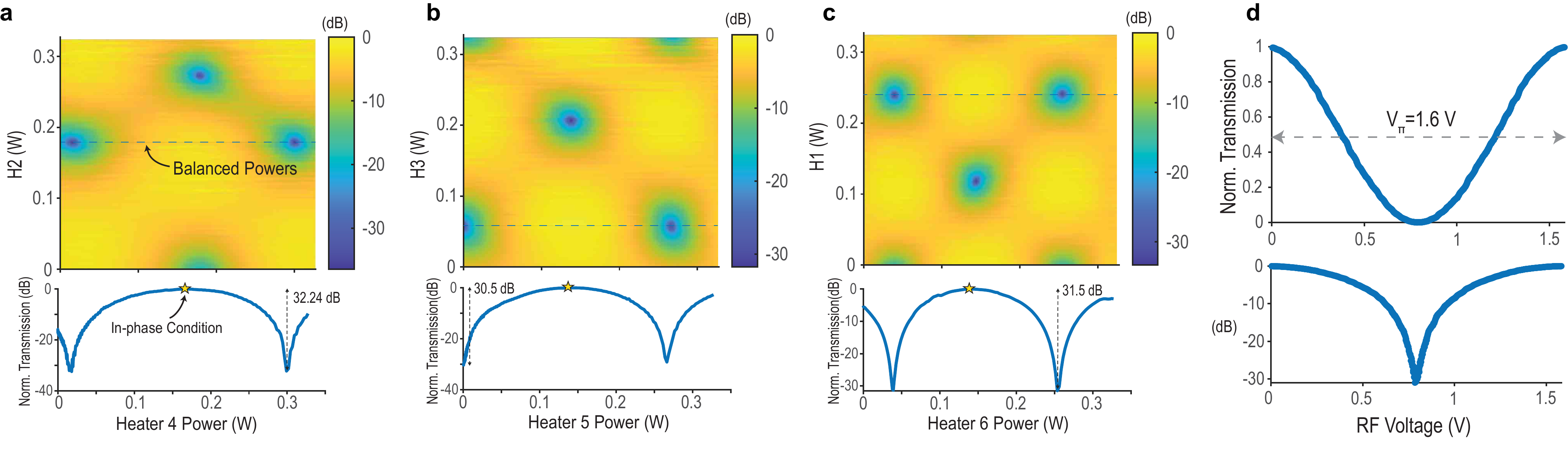}
             \begin{flushleft}
   Extended Data Figure 5.Characterization of integrated heaters and electro-optic half-wave voltage \textbf{a}-\textbf{c}. (Upper) 2D maps of normalized optical transmission with no RF in logarithmic scale as a function of heater electrical power for the upper channel pair (H2, H4), the lower channel pair (H3, H5), and the inter-pair (H1, H6), respectively. H1, H2, and H3 control power balance between the channels while H4, H5, and H6 control phase relationship between the channels. Dashed lines indicate the heater settings balancing optical powers between the optical channels. (Lower) Normalized optical transmission plotted in logarithmic scale as a function of the heater electrical power at the dashed lines, showing static extinction ratios of 32.24 dB (Ch1 and 2), 30.5 dB (Ch 3 and 4), and 31.5 dB (Ch1-4). The laser wavelength was set to approximately 790 nm. \textbf{d}. (Upper) Normalized optical transmission through a single push-pull pair as a function of low-frequency RF drive voltage, indicating a single-arm modulation half-wave voltage at $\approx$ 1.6 V. (Lower) Same data plotted on a logarithmic scale. 
        \end{flushleft}
  \label{fig:heater}
\end{figure*}

\clearpage  
\onecolumngrid

\section*{Supplementary Information}

\vspace{1em}
\begin{center}
\textbf{\large Integrated broadband optical isolator via dynamic rotating destructive interference}

\vspace{1em}
Kyunghun Han\textsuperscript{1,2,3,*}, Yiliang Bao\textsuperscript{1}, Junyeob Song\textsuperscript{1,2}, David Long\textsuperscript{1}, Sean Bresler\textsuperscript{1,4}, Daron Westly\textsuperscript{1}, Jason Gorman\textsuperscript{1}, Thomas LeBrun\textsuperscript{1}, Kartik Srinivasan\textsuperscript{1}, Vladimir Aksyuk\textsuperscript{1,†}

\vspace{1em}
\textsuperscript{1}Microsystems and Nanotechnology Division, National Institute of
Standards and Technology, 100 Bureau Drive, Gaithersburg, Maryland, 20899,
USA \\
\textsuperscript{2}Theiss Research, La Jolla, California, 92037, USA\\
\textsuperscript{3}Whiting School of Engineering, Johns Hopkins University, 3400 N. Charles Street, Baltimore, Maryland, 21218, USA\\
\textsuperscript{4}University of Maryland, College Park, Maryland, 20742, USA

\vspace{1em}
\textsuperscript{*}{kyunghun.han@nist.gov} ,
\textsuperscript{†}{vladimir.aksyuk@nist.gov}
\end{center}

\vspace{2em}
\subsection{Determination of the base RF frequency for maximizing forward transmission}
 \begin{figure}[htb!]
     \centering
     \includegraphics[width=1\linewidth]{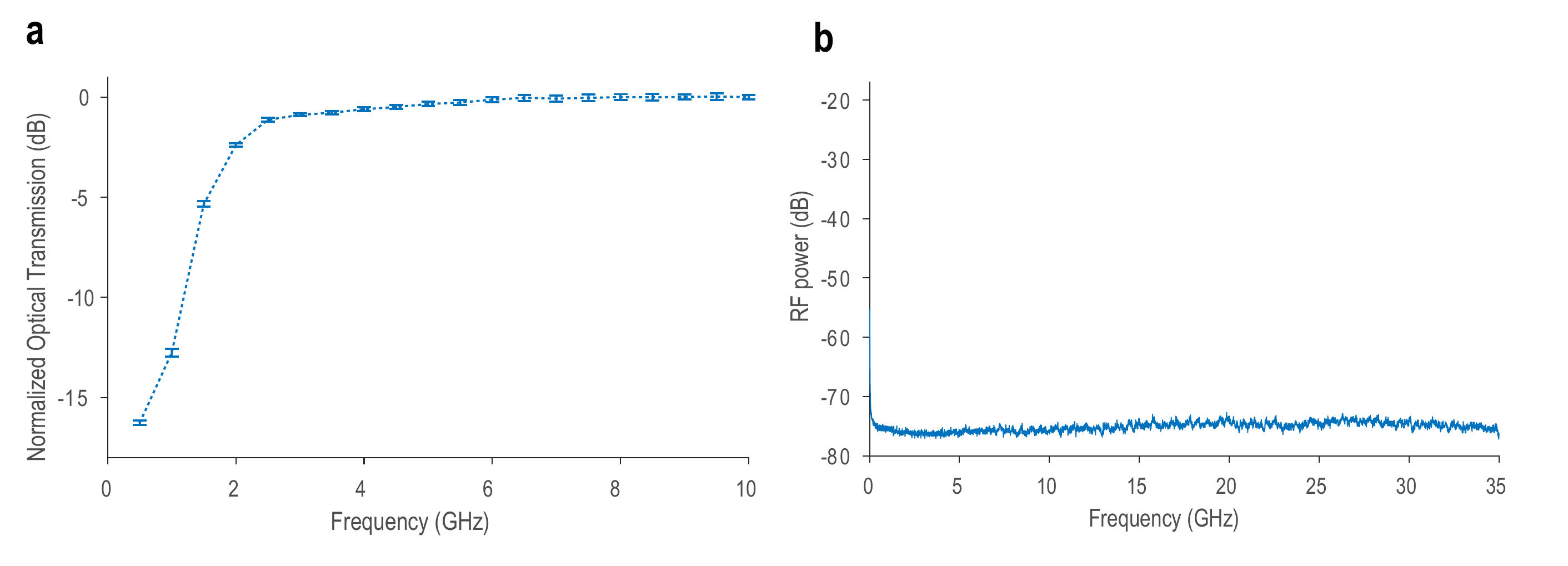}
         \begin{flushleft}
Figure S1. Frequency response of the forward propagating light \textbf{a}. Normalized optical transmission as a function of sinusoidal RF drive frequency. At 0.5 GHz, the transmission is strongly attenuated below -15 dB. As frequency increases, the zero mean phase modulation yields a reduced net phase shift and higher optical transmission. All values are normalized to the transmission at 10 GHz. \textbf{b}. RF power spectrum of the forward-propagating light. Measured resolution bandwidth is 5 MHz. The RF power scale is referenced to 1 mW. 
    \end{flushleft}    \label{fig1}
 \end{figure}
We swept the sinusoidal RF drive frequency from 0.5 GHz to 10 GHz while monitoring the forward propagating optical transmission. Figure S1a shows transmission is strongly attenuated at 0.5 GHz, indicating that the electro-optic modulator remains in a near electrostatic regime in which the device lengths are much shorter than the wavelength of the RF wave. As frequency increases, the net accumulated phase shift of the forward propagating light decreases because the periodic RF waveform has zero mean, leading to increased transmission. Above approximately 2.5 GHz, the slope begins to flatten, and the response saturates around 6.5 GHz. We therefore selected 6.5 GHz as the base RF frequency for driving, since higher frequencies give no further improvement in forward transmission. Figure S1b presents the RF spectrum of the forward propagating optical signal at a 6.5 GHz drive frequency, measured with a high frequency photoreceiver. The DC power level was approximately 32 µW, and the residual phase modulation sidebands stay below noise level.

 \subsection{RF and heater optimization }
 
  \begin{figure*}[htb!]
     \centering
     \includegraphics[width=1\linewidth]{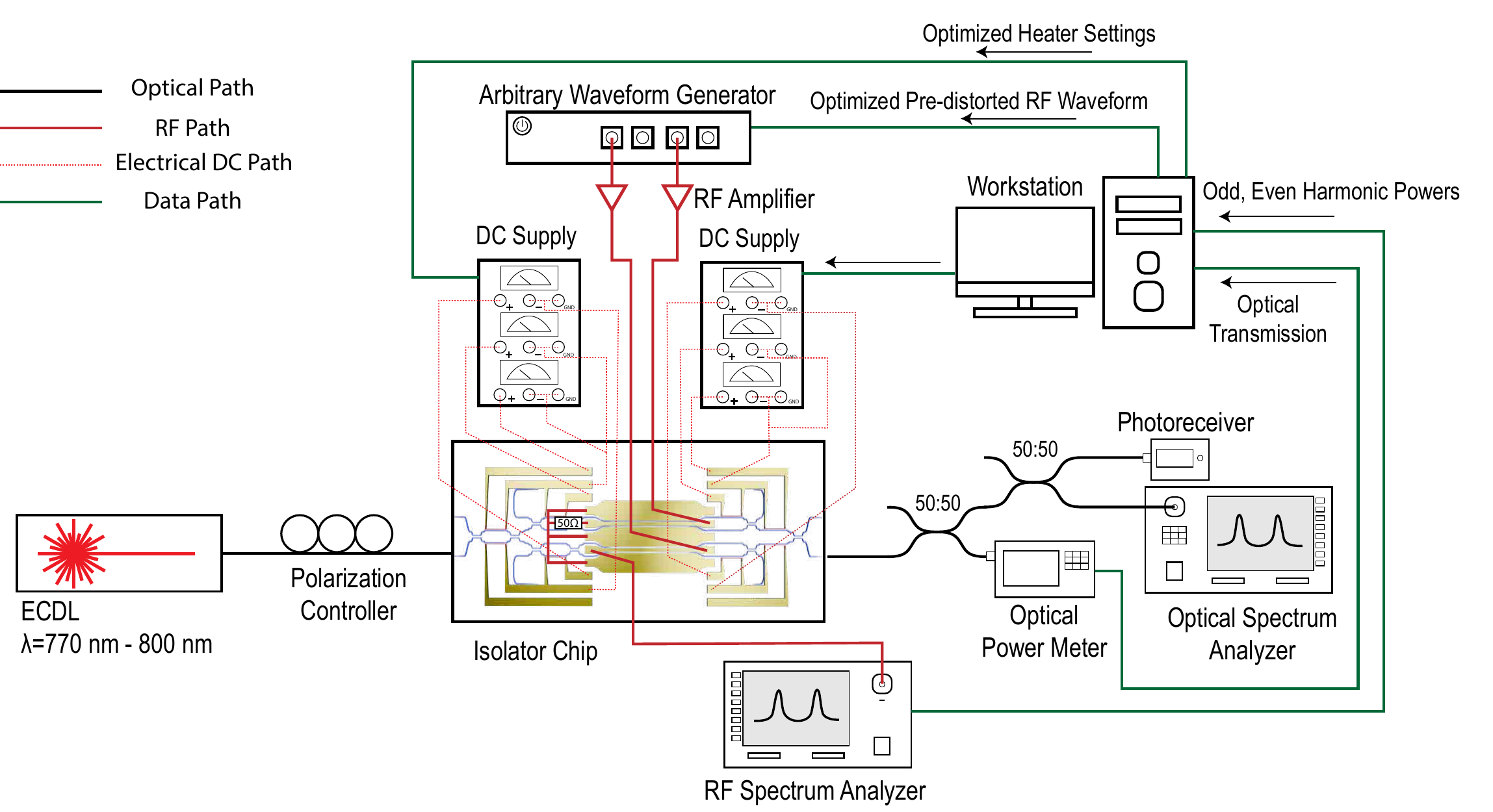}
         \begin{flushleft}
Figure S2. Experimental setup for the DRDI isolator characterization. The external-cavity diode laser is polarization-controlled and coupled into the isolator chip via tapered lensed fiber with a nominal mode field diameter of 2 µm. Multi-contact RF probes deliver two RF lines (solid red) to the CPW and six DC lines (dashed red) to on-chip microheaters. A two-channel AWG feeds pre-distorted RF wave. All instruments connect via data/control lines (green) to a central workstation, which iteratively adjusts the AWG waveform and heater settings.
    \end{flushleft}    \label{figS1}
 \end{figure*}

 The experimental setup comprises an external-cavity diode laser followed by an in-line optical fiber polarization controller, which sets the light to transverse-electric (TE) polarization at the isolator chip. Multi-contact RF probes, each carrying two RF and six DC lines, are placed at the CPW input and output. At the CPW output, the CPW may be terminated with a 50 $\Omega$ load or connected to an RF spectrum analyzer (also 50 $\Omega$), giving identical impedance conditions. An arbitrary waveform generator (AWG) with two independent RF channels produces multi-tone waveforms, which are amplified by RF amplifiers. Two three-channel DC power supplies drive the on-chip microheaters. The isolator’s output light is split via 50:50 directional couplers into three paths. An optical power meter measures time-averaged optical power. An optical spectrum analyzer detects any high-order harmonics. A high-frequency photoreceiver measures residual phase-modulations. All instruments are controlled by, and their data are acquired on a workstation.

   \begin{figure*}[htb!]
     \centering
     \includegraphics[width=0.5\linewidth]{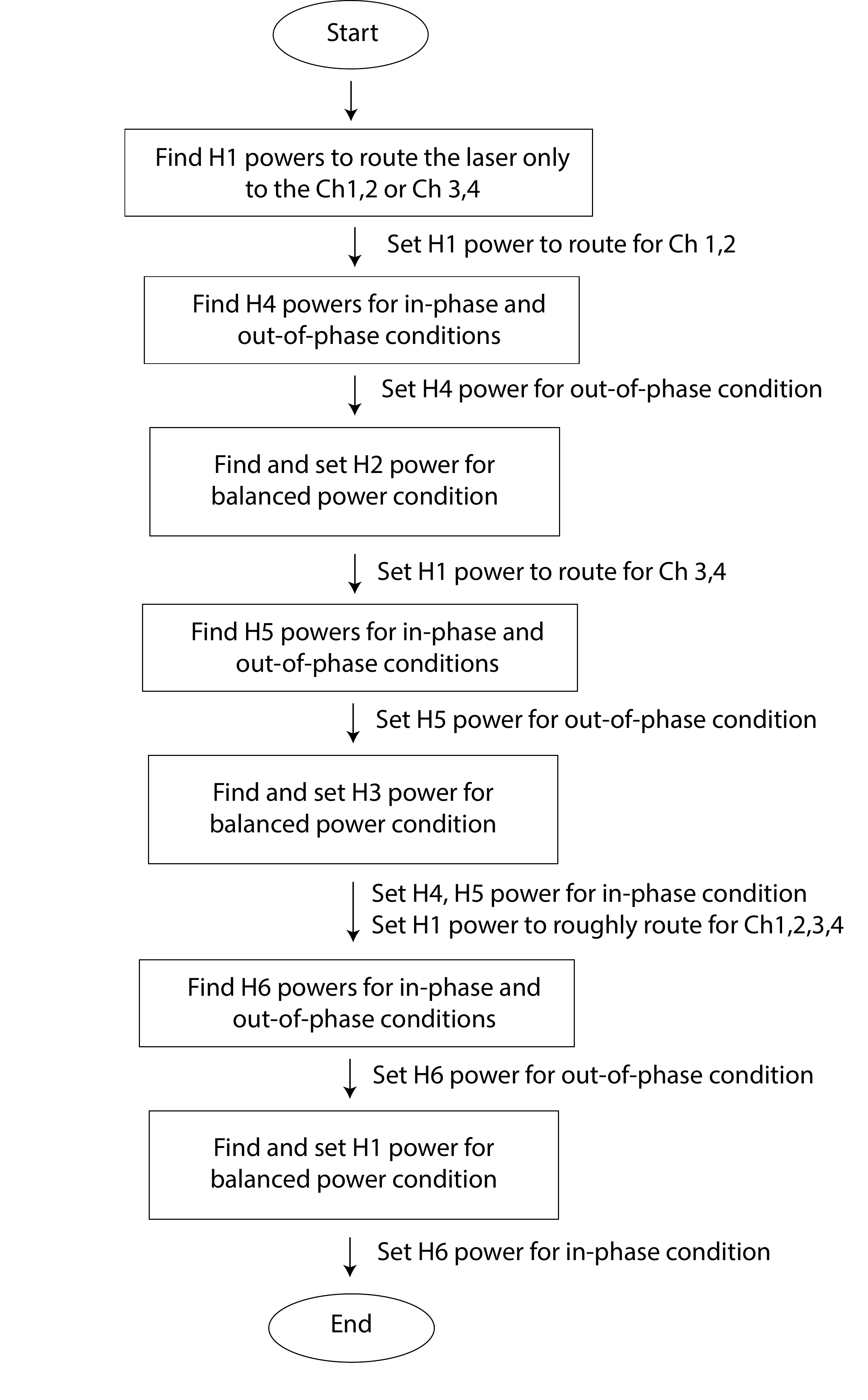}
         \begin{flushleft}
Figure S3. Flowchart for the microheater optimization. H1 first routes light into either the upper or lower push-pull pair. Each pair is set at the out-of-phase condition, then adjusted to balance power, and then switched to the in-phase condition. H6 is tuned to the out-of-phase condition to balance power between the pairs, then switched to the in-phase condition.
    \end{flushleft}    \label{figS2}
 \end{figure*}

 The microheaters in the isolator chip are driven by two DC power supplies controlled by a workstation. Because the split ratio of the parallel-waveguide directional couplers depends on wavelength, each operating wavelength requires its own optimized set of DC powers. First, we use an optical microscope to observe scattered light and adjust Heater 1 (H1) to route all input laser power into either the upper push-pull pair (Ch 1 and 2) or the lower push-pull pair (Ch 3 and 4). With H1 fixed for the upper pair, we sweep heater 4 (H4) and measure the output optical power. The transmission minimum marks the out-of-phase condition, and the maximum marks the in-phase condition. The out-of-phase point is easier to identify since its transmission change exceeds noise fluctuations. To locate the in-phase point accurately, we fit the transmission data to a sinusoidal function to suppress noise. Next, holding H4 at its out-of-phase setting, we sweep heater 2 (H2) to balance power between Ch1 and Ch2, again indicated by a transmission minimum. We then set H1 to direct most power into the lower pair and repeat the optimization for that pair by sweeping heater 5 (H5) and 3 (H3) to minimize transmission. After both pairs are optimized, we set H4 and H5 to their in-phase conditions to establish the dynamic rotating destructive interference (DRDI). Finally, we sweep heater 6 (H6) to find its out-of-phase point, use that to fine-tune H1 for equal splitting between the two pairs (minimum transmission), and then set H6 to its in-phase point for the final DRDI operation. 

    \begin{figure}[htb!]
     \centering
     \includegraphics[width=0.5\linewidth]{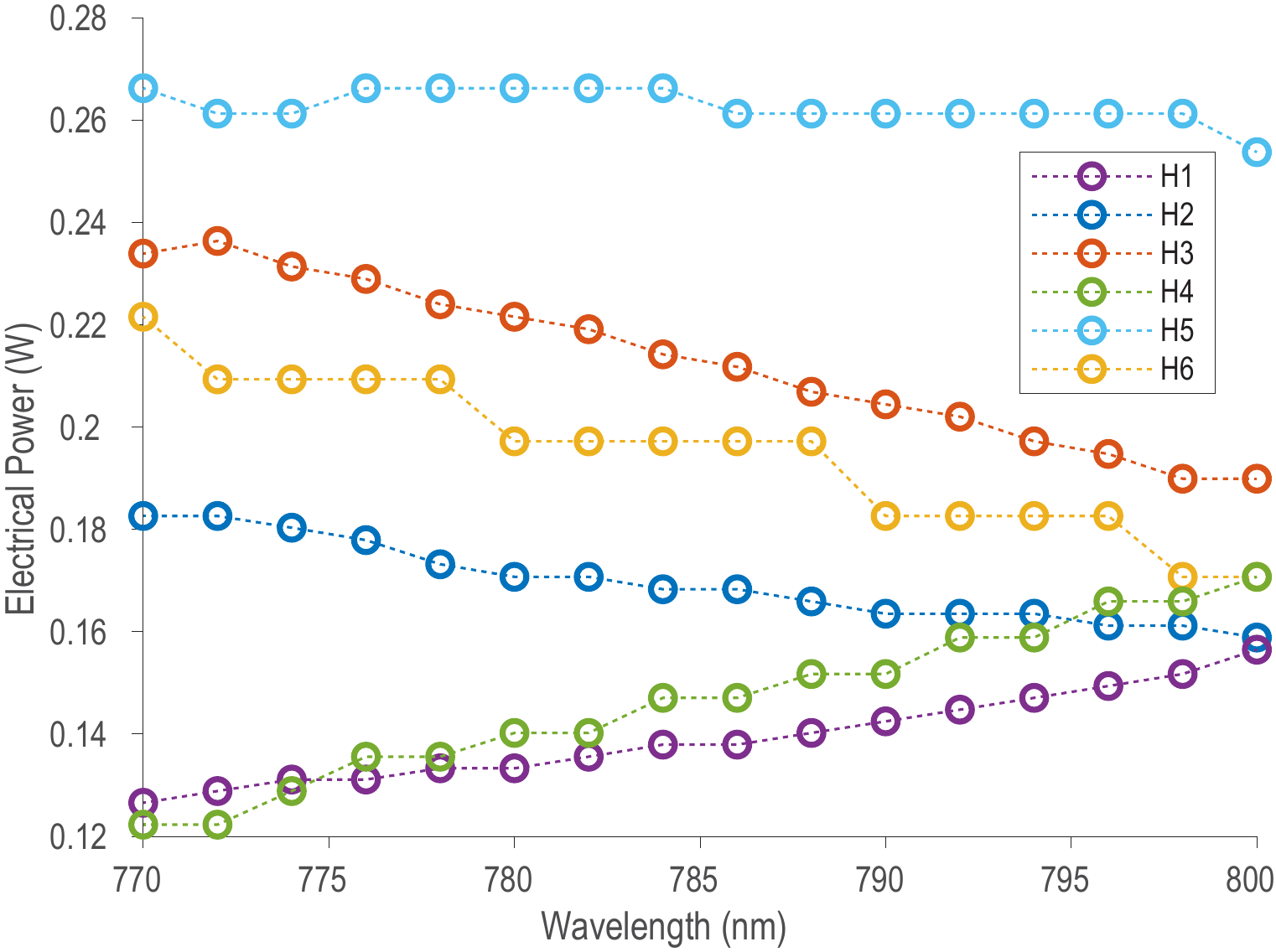}
         \begin{flushleft}
Figure S4. Optimized heater powers for various wavelengths. Electrical powers for H1-H6 are plotted as a function of the laser wavelengths. All heaters exhibit linear tuning across the wavelength range.
    \end{flushleft}    \label{figS3}
 \end{figure}
 
 The optimized heater powers at each wavelength change almost linearly across the wavelength range from 770 nm to 800 nm as shown in Figure S4. The discrete steps on H6 arise from the coarse scanning increments used in the optimization for faster optimization. Importantly, the heater settings vary continuously without jumps, enabling faster and more stable wavelength sweeps. This can be advantageous for any spectroscopic applications with sweeping ECDL wavelength. We confirmed that measuring only a few wavelengths and interpolating the heater powers between them gives nearly identical isolation to the method of optimizing heaters at every wavelength. 

    \begin{figure}[htb!]
     \centering
     \includegraphics[width=0.3\linewidth]{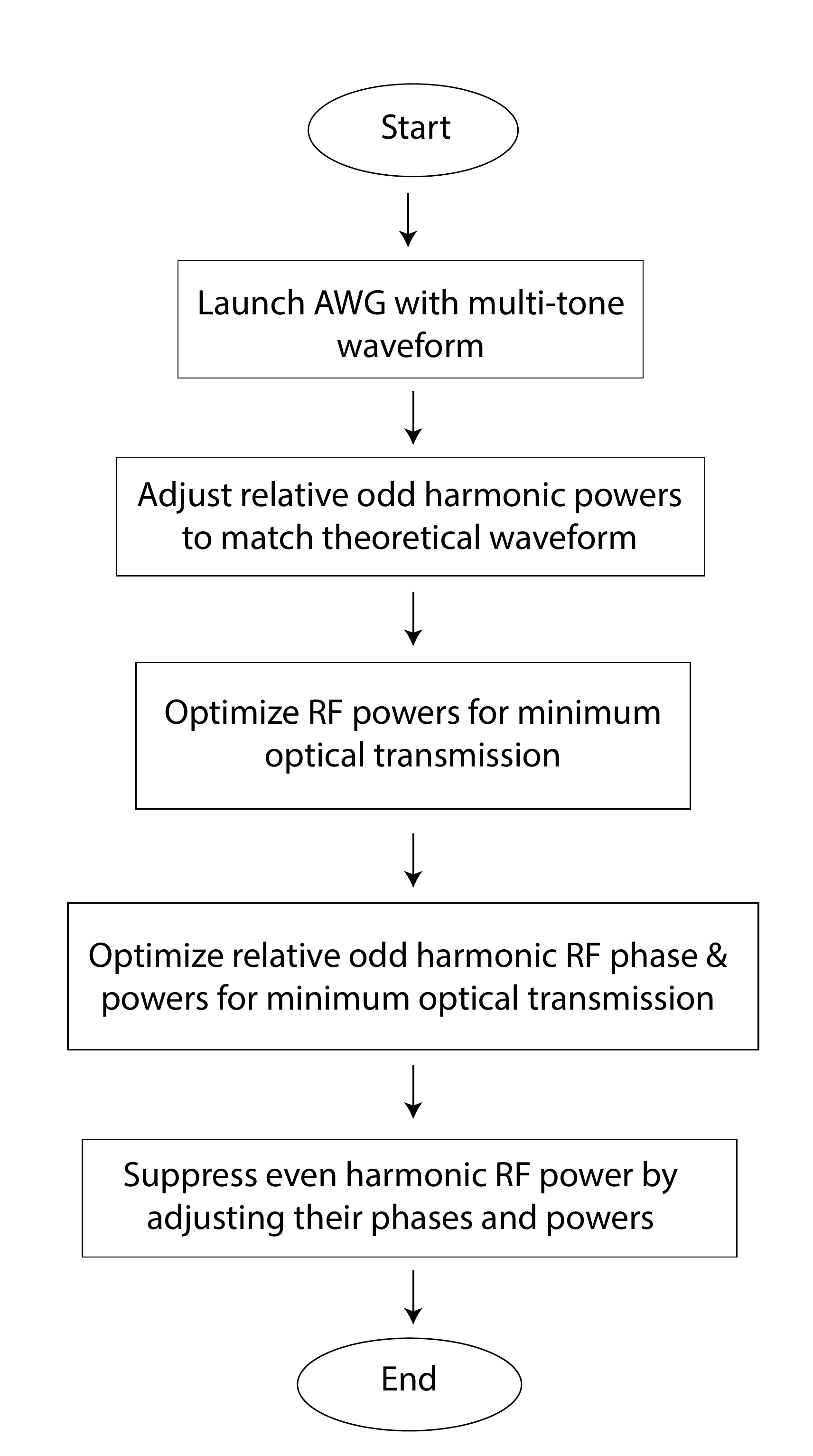}
         \begin{flushleft}
Figure S5. Flow chart for RF waveform optimization. To overcome nonlinearity and attenuation in the RF path, we generated a multi-tone RF waveform consisting of the fundamental and odd harmonics and iteratively adjusted their amplitudes and phases based on the measured CPW output spectrum and optical transmission minima. Any residual even harmonics are suppressed by fine tuning their phases and powers.     \end{flushleft}    \label{figS4}
 \end{figure}

 Because of nonlinearity, frequency-dependent phase delay, and attenuation, we must optimize the RF drive waveform to maximize isolation. We generate the multi-tone waveform that its degree of freedom are the magnitudes and phases of the fundamental and its odd harmonics. Truncating the harmonics to only odd harmonics simplifies the optimization by eliminating redundant parameters required by the DRDI scheme. The detailed procedure is shown in Figure S5. First, the AWG outputs the theoretically predicted waveform, and we compare the measured RF spectrum at the CPW output with the ideal spectrum to adjust the relative powers of the high-order odd harmonics. Next, for each push-pull pair, we sweep the AWG output power to identify the setting that minimizes the backward optical transmission. We then optimize each odd harmonic by adjusting their phases and magnitudes individually. Because even harmonics are unnecessary for DRDI, we adjust their phases and amplitudes to compensate for any unwanted even harmonic generation by the RF amplifiers. We repeat the entire cycle until the target isolation is reached.

 \subsection{RF waveform requirement for DRDI}

      \begin{figure}[htb!]
     \centering
     \includegraphics[width=1\linewidth]{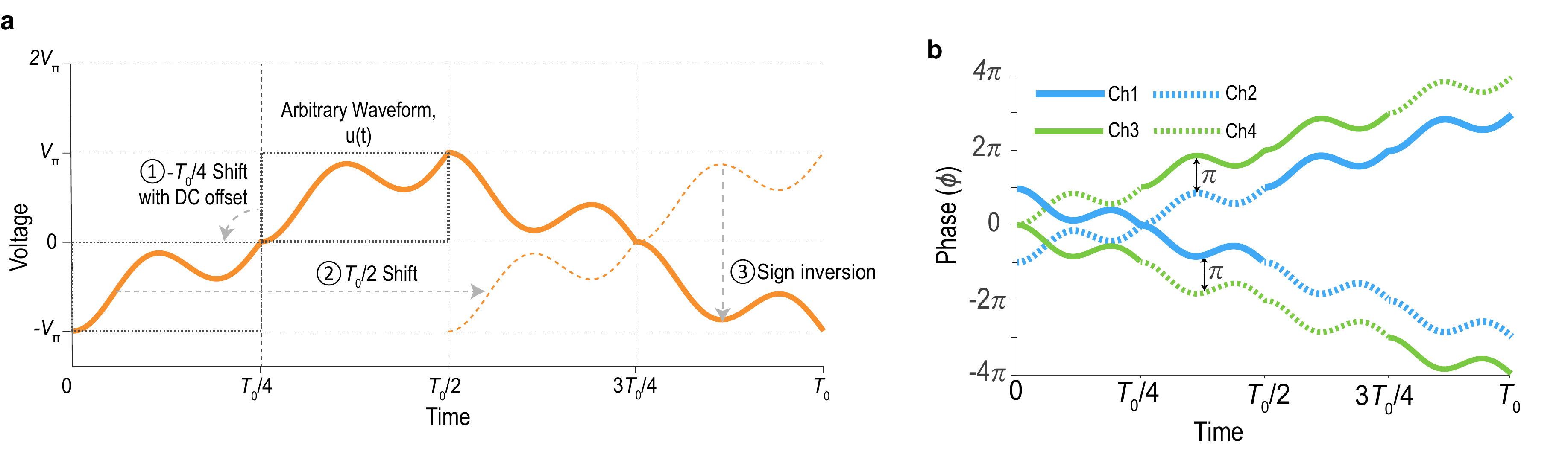}
         \begin{flushleft}
Figure S6. Generalized RF waveform synthesis and resulting phases for the DRDI. \textbf{a}. Arbitrary base waveform $u(t)$ defined on the quarter-period interval is extended over the full period. The solid line shows the resulting period, zero-mean waveform enforcing half-wave symmetry \textbf{b}. Time-segmented, unwrapped accumulated phases of the four channels over one RF period ($T_0$). At each quarter-period, the channels form two cancellation pairs, each maintaining a $\pi$ phase difference, implementing the DRDI scheme.    \end{flushleft}    \label{figS5}
 \end{figure}

 Although a triangular waveform is a valid solution for DRDI, it is impractical because truncating its harmonics dramatically degrades isolation. To find an optimal waveform within a given RF bandwidth, we generalize the required RF waveform described in Figure S6a. First, the RF waveform must be periodic and zero‑mean, so that the forward propagating optical wave experiences no net phase shift in the traveling‑wave modulator. For destructive interference between four optical waves driven by two RF signals delayed by one‑quarter RF period, we begin with a base RF voltage waveform \(u(t)\) defined on the quarter‑period interval from \(T_0/4\) to \(T_0/2\). The boundary conditions of the base waveform are

\begin{equation}\tag{S1}
u\!\bigl(\tfrac{T_0}{4}\bigr) = 0,\quad
u\!\bigl(\tfrac{T_0}{2}\bigr) = V_{\pi}
\end{equation}

Then we can extend the RF voltage waveform \(f(t)\) to all \(t\) by enforcing periodicity \(f(t-T_0)=f(t)\) and half‑wave symmetry \(f(t-T_0/2)=-f(t)\). This can be achieved for the period \(0 \le t \le T_0\) with the conditions

\begin{equation}\tag{S2}
f(t) =
\begin{cases}
u(t+T_0/4) - V_{\pi}, & 0 \le t < T_0/4,\\[6pt]
u(t), & T_0/4 \le t < T_0/2,\\[6pt]
-\,f\bigl(t - T_0/2\bigr), & T_0/2 \le t < T_0.
\end{cases}
\end{equation}

Because of Equation (S1), these conditions guarantee continuity at the quarter‑period boundaries, while automatically satisfying half‑wave symmetry.

\end{document}